\documentclass[preprint,12pt,authoryear]{elsarticle}

\usepackage{epstopdf} 

\usepackage{amssymb}
\usepackage{amsthm}
\usepackage[intlimits]{amsmath}

\usepackage[latin1]{inputenc} 
\usepackage{caption}
\usepackage{subcaption}
\usepackage{booktabs}
\usepackage{lscape}

\newcommand{\E}{{\mathbb E}}

\renewcommand{\P}{{\mathbb P}}

\newcommand{\U}{{\mathbb U}}
\newcommand{\V}{{\mathbb V}}
\newcommand{\W}{{\mathbb W}}

\newcommand{\Z}{{\mathbb Z}}

\newcommand{\BB}{{\mathcal{B}}}
\newcommand{\DD}{{\mathcal{D}}}
\newcommand{\EE}{{\mathcal{E}}}

\newcommand{\KK}{{\mathcal{K}}}

\newcommand{\OO}{{\mathcal{O}}}
\newcommand{\PP}{{\mathcal{P}}}

\newcommand{\UU}{{\mathcal{U}}}
\newcommand{\VV}{{\mathcal{V}}}
\newcommand{\WW}{{\mathcal{W}}}

\newcommand{\bsa}{\boldsymbol a}
\newcommand{\bsb}{\boldsymbol b}

\newcommand{\bsP}{\boldsymbol P}

\newcommand{\bsW}{\boldsymbol W}

\newcommand{\bsnull}{\boldsymbol 0}

\newcommand{\bsxi}{\boldsymbol \xi}

\newcommand{\bsomega}{\boldsymbol \omega}

\newcommand{\bsOmega}{\boldsymbol \Omega}

\newcommand{\bsXi}{\boldsymbol \Xi}

\newcommand{\eps}{{\varepsilon}}

\DeclareMathOperator*{\argmin}{arg\,min}

\DeclareMathOperator{\var}{\V ar}

\newcommand{\ov}\overline
\newcommand{\what}{\widehat}
\newcommand{\wtilde}{\widetilde}
\newcommand{\ow}{\text{ otherwise}}
\newcommand{\rig}\right
\newcommand{\lef}\left
\newcommand{\nf}\normalfont

\journal{Applied Energy}

\graphicspath{{Pics/}}

\begin{document}

\begin{frontmatter}

\title{Forecasting Wind Power -- Modeling Periodic and Non-linear
Effects Under Conditional He\-te\-ro\-sce\-das\-ti\-ci\-ty}

\author{Florian Ziel}
\address{European University Via\-dri\-na, Chair for Finance and Capital Market Theory, Große Scharrnstraße 59, 15230 Frankfurt (Oder), Germany, Tel. +49 (0)335 5534 2986, E-Mail: ziel@europa-uni.de.}

\author{Carsten Croonenbroeck}
\address{University of Rostock, Faculty for Environmental Science, Justus-von-Liebig-Weg 2, 18059 Rostock, Germany, Tel. +49 (0)381 498 3267, E-Mail: carsten.croonenbroeck@uni-rostock.de.}

\author{Daniel Ambach}
\address{European University Via\-dri\-na, Chair for Quantitative Methods, esp. Statistics, Große Scharrnstraße 59, 15230 Frankfurt (Oder), Germany, Tel. + 49 (0)335 5534 2983, E-Mail: ambach@europa-uni.de.}

\begin{abstract}
\footnotesize{In this article we present an approach that enables
joint wind speed and wind power forecasts for a wind park. We
combine a multivariate seasonal time varying threshold
autoregressive moving average (TVARMA) model with a power threshold
generalized autoregressive conditional heteroscedastic
(power-TGARCH) model. The modeling framework incorporates diurnal
and annual periodicity modeling by periodic B-splines, conditional
he\-te\-ro\-sce\-das\-ti\-ci\-ty and a complex autoregressive
structure with non-linear impacts. In contrast to usually
time-consuming estimation approaches as likelihood estimation, we
apply a high-dimensional shrinkage technique. We utilize an
iteratively re-weighted least absolute shrinkage and selection
operator (lasso) technique. It allows for conditional
he\-te\-ro\-sce\-das\-ti\-ci\-ty, provides fast computing times and
guarantees a parsimonious and regularized specification, even though
the parameter space may be vast. We are able to show that our
approach provides accurate forecasts of wind power at a
turbine-specific level for forecasting horizons of up to 48 hours
(short- to medium-term forecasts).}
\end{abstract}


\end{frontmatter}


\clearpage

\section{Introduction}\label{section:Introduction}

Wind power is on the verge of becoming the most important source of electricity in many countries worldwide. \cite{berkhout2013wind} argue that wind power is the most emergent renewable power source with a growth rate of $30 \%$ per year. However, the technology still has a few challenges to master. In contrast to conventional power, wind power production is non-deterministic and highly volatile. To make efficient contracts at the energy pools, accurate forecasts of wind power production have to be available. \cite{Lei2009} as well as \cite{Soman2010} provide a time-scale classification of wind power and wind speed prediction models. Longer-term forecasts at horizons of two days up to one week for, e.g., decisions with respect to the required energy reserves and the maintenance scheduling, are based on meteorological or recently developed hybrid structure models.\\
Match-making at the energy markets, i.e. trading at the usual
day-ahead markets common to most energy pools, requires predictions
at forecasting horizons of up to 48 hours, at most, dependent on the
designated contract market. Forecasts for this medium- to long-term
scenario are usually based on stochastic modeling, on artificial
intelligence models or specific neural networks. For instance,
\cite{cadenas2009short}, \cite{cao2012forecasting} and
\cite{azad2014long} use them. \cite{amjady2011short} use ridgelet
neural networks which possess ridge functions as activators for
their hidden nodes to provide forecasts of the aggregated wind power
output of a wind farm. \cite{bhaskar2012awnn} take a statistical
approach which does not use numerical weather predictions. They use
a wavelet decomposition of their wind speed time series and an
adaptive wavelet neural network. After transformation, they transfer
the wind speed predictions by using a feed-forward neural network
into wind power forecasts. \cite{liu2014short} propose a hybrid
model which combines inputs selected by deep quantitative analysis,
wavelet transform, genetic algorithm and support vector machines.
Another wavelet support vector machine approach is used by
\cite{zeng2012short} to perform wind power predictions.
\cite{zhou2013application} apply a probabilistic kernel density
forecasting model with a quantile-copula estimator to perform wind
power forecasts. They evaluate the
model by using a power system in Illinois and compare several scheduling strategies. \cite{haque2014hybrid} provide a new hybrid intelligent algorithm for wind power predictions that uses a combination of wavelet transform and fuzzy network methods.\\
For match-making, stochastic forecasting approaches like the one
presented in this paper benefit from modeling the persistence of
wind power, its periodic structure and its direct dependence on wind
speed. Thus, wind speed itself is usually predicted in an entirely
stochastic setting, while also, numerical weather predictions (NWPs)
can be employed, if available. Powerful statistical models return reliable forecasts of wind power for short- to medium-term scenarios and are widely established, like the Wind Power Prediction Tool (WPPT) by \cite{Nielsen2007}, its recent generalization, \mbox{GWPPT}, by \cite{Croonenbroeck2014}, or the spatial \mbox{GWPPT} by \cite{Croonenbroeck2015613}. However, the class of statistical approaches also incorporates autoregressive (AR), autoregressive moving average (ARMA) and autoregressive fractionally integrated moving average (ARFIMA) models. \cite{kavasseri2009day} discuss these models in details.\\
The literature on contributed models for wind power and wind speed
forecasts is vast, \cite{Jung2014} as well as
\cite{Tascikaraoglu2014} provide an up-to-date overview. Most models
have several drawbacks: One problem is that stochastic wind power
prediction models require wind speed forecasts in the first step.
Several models for wind speed forecasting are available, as provided
by \cite{zhu2014space}, \cite{ambach2015periodic} or
\cite{shukur2015daily}. In the second step, these predictions are
transformed into forecasts of wind power, as shown by, e.g.,
 \cite{azad2014long}. Most of the models do not provide
conjoint wind power and wind speed predictions. Many models, e.g.
the aforementioned WPPT class models, utilize wind speed as a
quadratic regressor for wind power, although the theoretical
non-linear relationship is usually described by a cubic function.
The reason for this is to be found in the physical limitation of the
turbine, i.e. the upper bound of producible wind power.\\
The long memory structure of usual turbine specific wind speed and
wind power data suggests a diurnal and an annual periodic behavior.
Several contributions illustrate this periodic or cyclic behavior, as, e.g., \cite{carapellucci2013effect}, \cite{silva2016complementarity}, \cite{scholz2014cyclic} and \cite{ambach2015periodic}. \cite{ambach2015short} focuses on annual periodic effects.\\
Periodic effects may change over time, which is usually not
considered. \cite{ambach2015short} and \cite{Ambach2015Space}
incorporate seasonal interactions in wind speed time series by
annual and diurnal basis functions. Thereby, they capture the annual
change of a daily period. This effect is basically driven by the
fact that the length of the nights changes over the year: On the
northern hemisphere, there are longer nights during the winter than
during the summer. Indeed, it is observable that the diurnal
periodicity is varying over the year.
Moreover, evidence suggests that such periodicities are also observable within the wind power data (see, e.g., \citealp{Nielsen2007}).\\
Consequently, our new forecasting model includes all common
stochastic modeling features, but in addition, it overcomes the
aforementioned drawbacks. The main advantage of our approach is
related to the fact that we are able to produce wind speed and power
forecasts at the same time with one model. The periodic behavior of
the day and the year is modeled by periodic B-splines. Furthermore,
we capture interaction between both seasonalities, as the diurnal
impact may change over the year. Thus, we allow for
periodic changes in the parameters to capture the seasonal interaction effects.\\
Wind speed and wind power show a huge amount of autocorrelation, as
shown by \cite{ambach2015periodic}. Hence, we consider a
multivariate seasonal VARMA class model to capture the persistence
as well as the periodicity. A VARMA model is also used by
\cite{erdem2011arma} to predict a tuple of wind speed time series.
In a more general setting, \cite{jeon2012using} take a bivariate
VARMA generalized autoregressive conditional heteroscedastic (GARCH)
approach to model the wind speed and wind direction and convert the
predictions of both into wind power forecasts.\\
Instead of using wind speed as a quadratic regressor as done in the
WPPT and \mbox{GWPPT} approach, we use thresholds and vector
autoregression to cover the non-linearity.
The threshold autoregressive approach is also applied in a context of probabilistic load forecasting by \cite{ziel2016lasso} as a suitable tool to explore the non-linearity in the data. Here, we use a VARMA model to capture the correlation structure of several turbines and to predict wind speed and wind power altogether. Finally, we propose a threshold GARCH (TGARCH) model for the wind speed series and a power-TGARCH process for the volatility. With it, we are able to capture the conditionally heteroscedastic behavior in the data, similarly to \cite{ewing2006time} and \cite{ambach2015periodic}.\\
The assumed statistical model structure for the wind speed and power
allows us to simulate sample paths for several scenarios. Using
bootstrap simulation techniques we can easily derive probabilistic
forecasts. As pointed out by, e.g., \cite{pinson2013wind},
\cite{alessandrini2013comparison}, and \cite{hong2016probabilistic},
the importance of probabilistic wind power forecasting is
increasing, especially for longer forecasting horizons.
\cite{Zugno2012} use probabilistic forecasts of wind power as well,
\cite{Gneiting2007} provide details on the computation and
evaluation of probabilistic forecasts. Recently, \cite{Berner2015}
discuss bias correction and accuracy improvements in general
probabilistic forecasting. Using a mesoscale meteorology framework,
they address the main reason for using probabilistic forecasts
instead of point forecasts, i.e. ``[... to] account for certain
aspects of structural model uncertainty''.\\
For the estimation, we apply a high-dimensional shrinkage technique
based on the popular least absolute shrinkage and selection operator
(lasso) method, as introduced by \cite{tibshirani1996regression}.
Similarly, \cite{evans2014} use the lasso method to augment the
forecasting accuracy of a wind farm. According to
\cite{ziel2015efficient}, we apply an iteratively re-weighted lasso
approach to estimate the model parameters. Thus, we can provide a
huge parameter space, still come up with a parsimonious and
regularized specification and have very convenient computing times
in comparison to the usual maximum likelihood technique (i.e. few
seconds compared to several minutes on a modern computer). For the
time varying and periodic effects, the algorithm will estimate
parameters that may vary over time at a certain significance.
Otherwise, the parameters remain constant. Our time varying periodic
TVARMA-power-TGARCH model returns more accurate forecasts than the
usual WPPT and \mbox{GWPPT} models as well as a set of benchmark models, including the usual persistence forecaster. Results show that our model provides less skewed forecast errors than our benchmarks.\\
This paper makes two major contributions: First, we present a modeling framework for wind power that includes wind speed, flexible modeling of the periodicity and he\-te\-ro\-sce\-das\-ti\-ci\-ty. Second, we show how to estimate the model parameters by applying a re-weighted he\-te\-ro\-sce\-das\-tic lasso approach to a time series setting, as has been done recently by \cite{ziel2015iteratively}. Empirical results from out-of-sample forecasts are compared to a set of benchmark models.\\
The paper is structured as follows: Section \ref{section:Data}
discusses the data set used. In Section \ref{section:Model} we show
our new model idea. Section \ref{section:Estimation} presents the
estimation technique. Empirical results are discussed in Section
\ref{section:Results} and Section \ref{section:Conclusion}
concludes.

\section{Data and Their Characteristics}\label{section:Data}

\begin{figure}[htb]
  \centering
 \includegraphics[width=.6\textwidth]{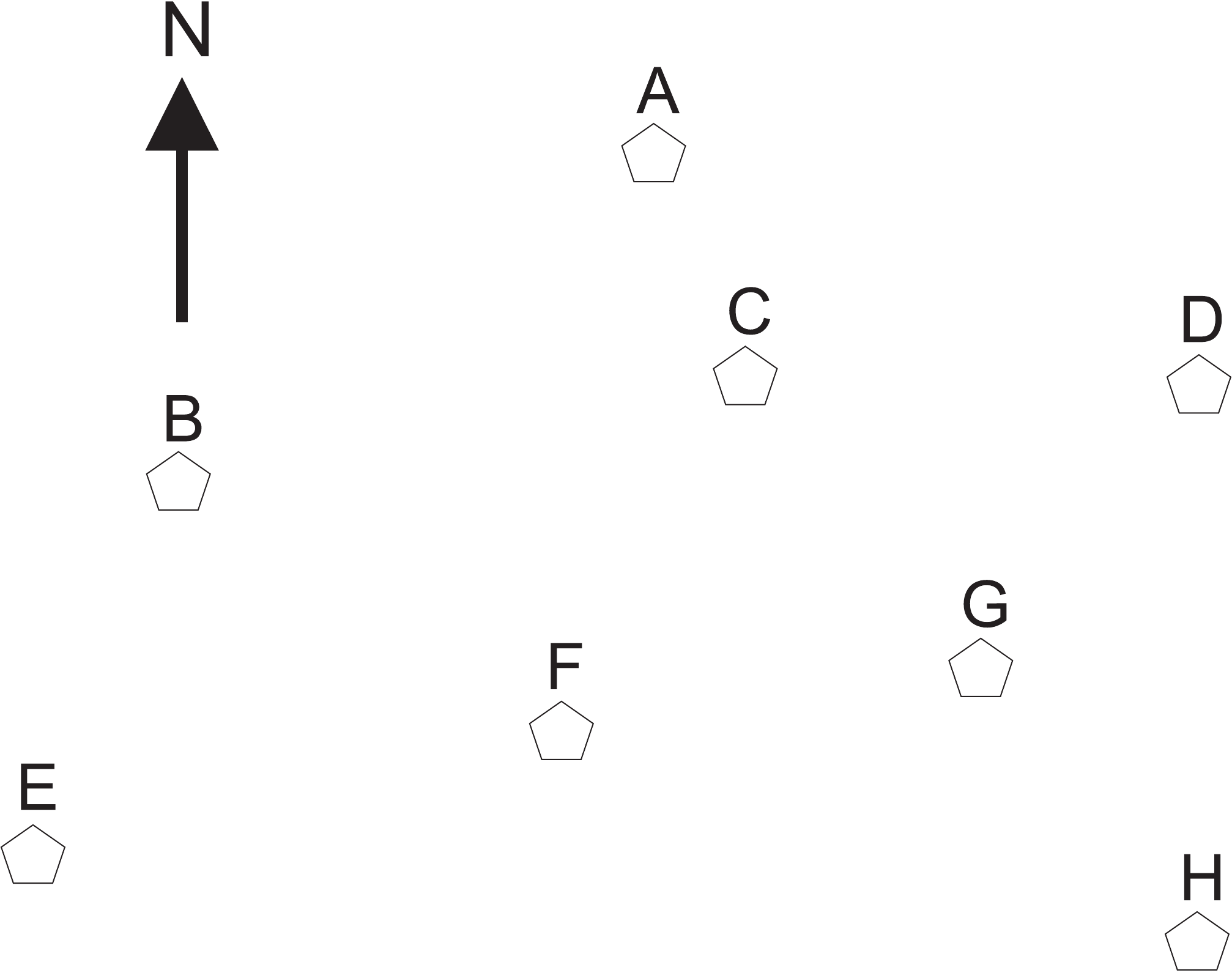}
  \caption{Stylized map of the wind parks investigated.}
  \label{graph:Map}
\end{figure}

The turbine data set used in this paper is a high-frequency series
collected from a wind park in Germany. The wind park consists of $8$
turbines. The observed park is situated in a mostly plain and rural
region. The area has a slight roughness with fields and some
forestation. Due to a non-disclosure agreement, the specific
locations cannot be revealed. However, Figure \ref{graph:Map}
presents a stylized map of the turbines' arrangement. The turbines,
labeled Turbine A to H, exhibit a power range of $[0;1500]$ kW each
and write sensor data to log files at a frequency of ten minutes.
The observed time frame spans from November 1, 2010 to November 5,
2012, so there are 105984 observations per turbine.

\begin{figure}[hbt!]
\centering
 \begin{subfigure}[b]{0.49\textwidth}
 \includegraphics[width=1\textwidth]{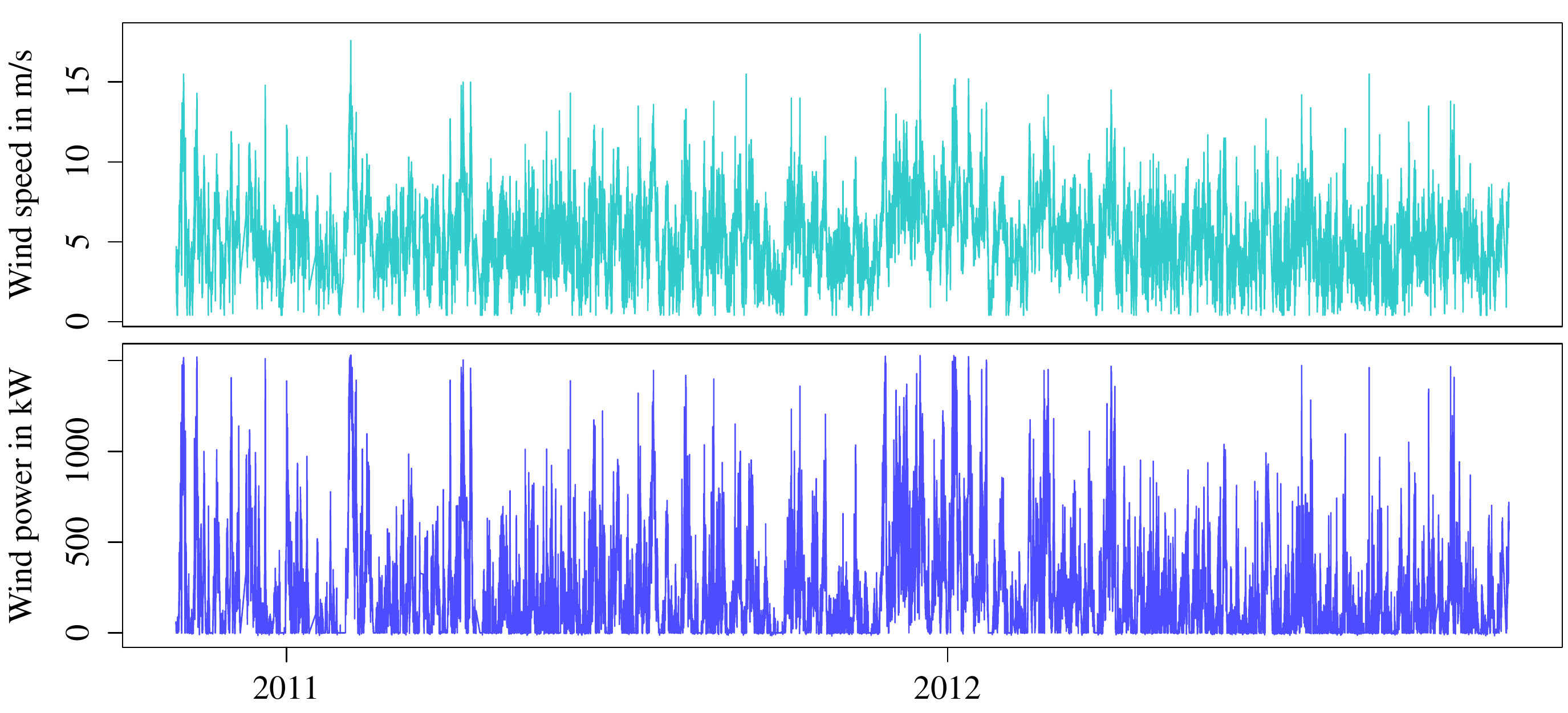}
 \includegraphics[width=.49\textwidth]{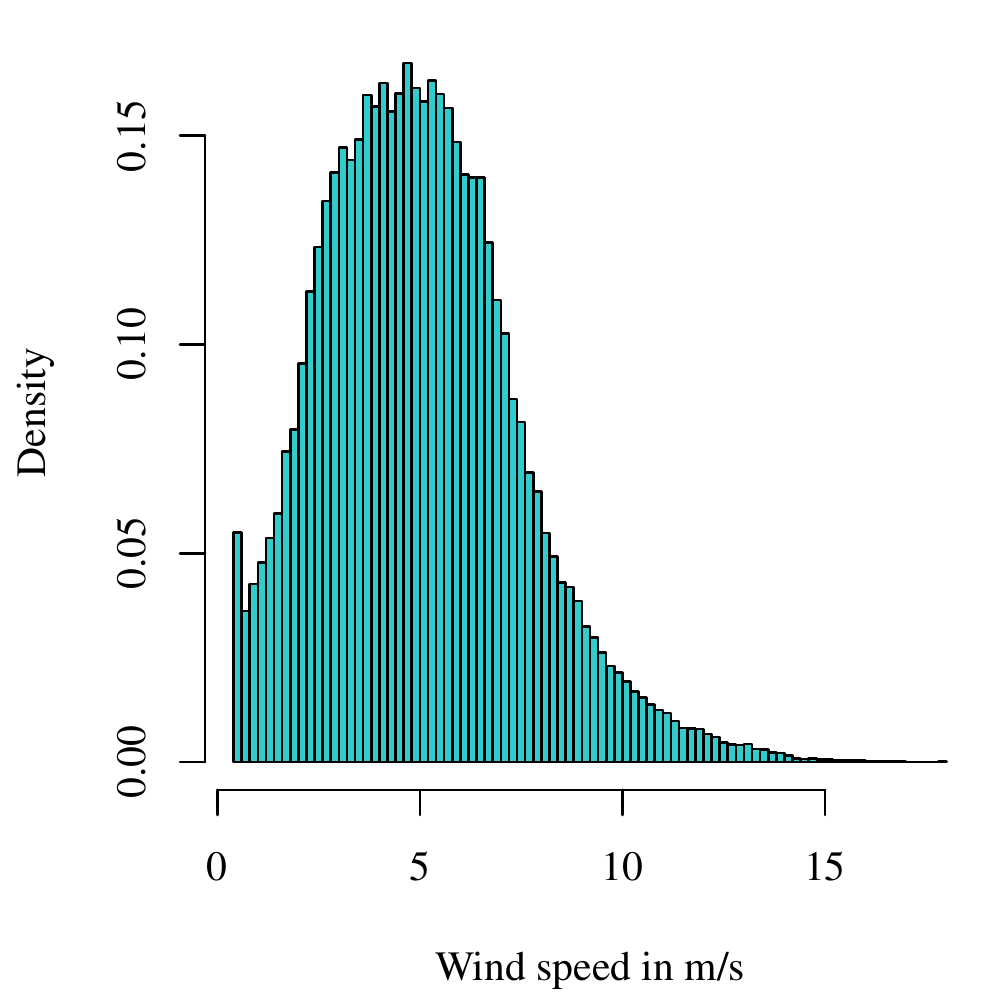}
 \includegraphics[width=.49\textwidth]{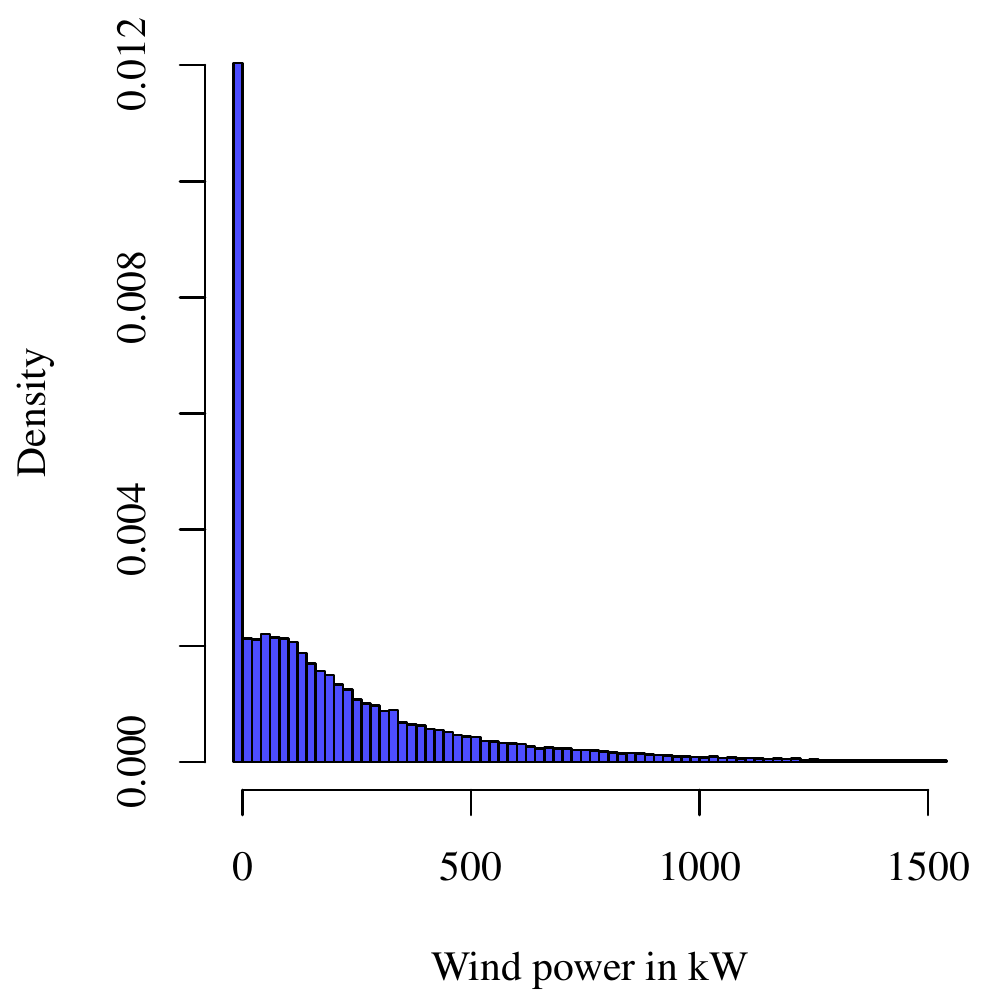}
 \includegraphics[width=1\textwidth]{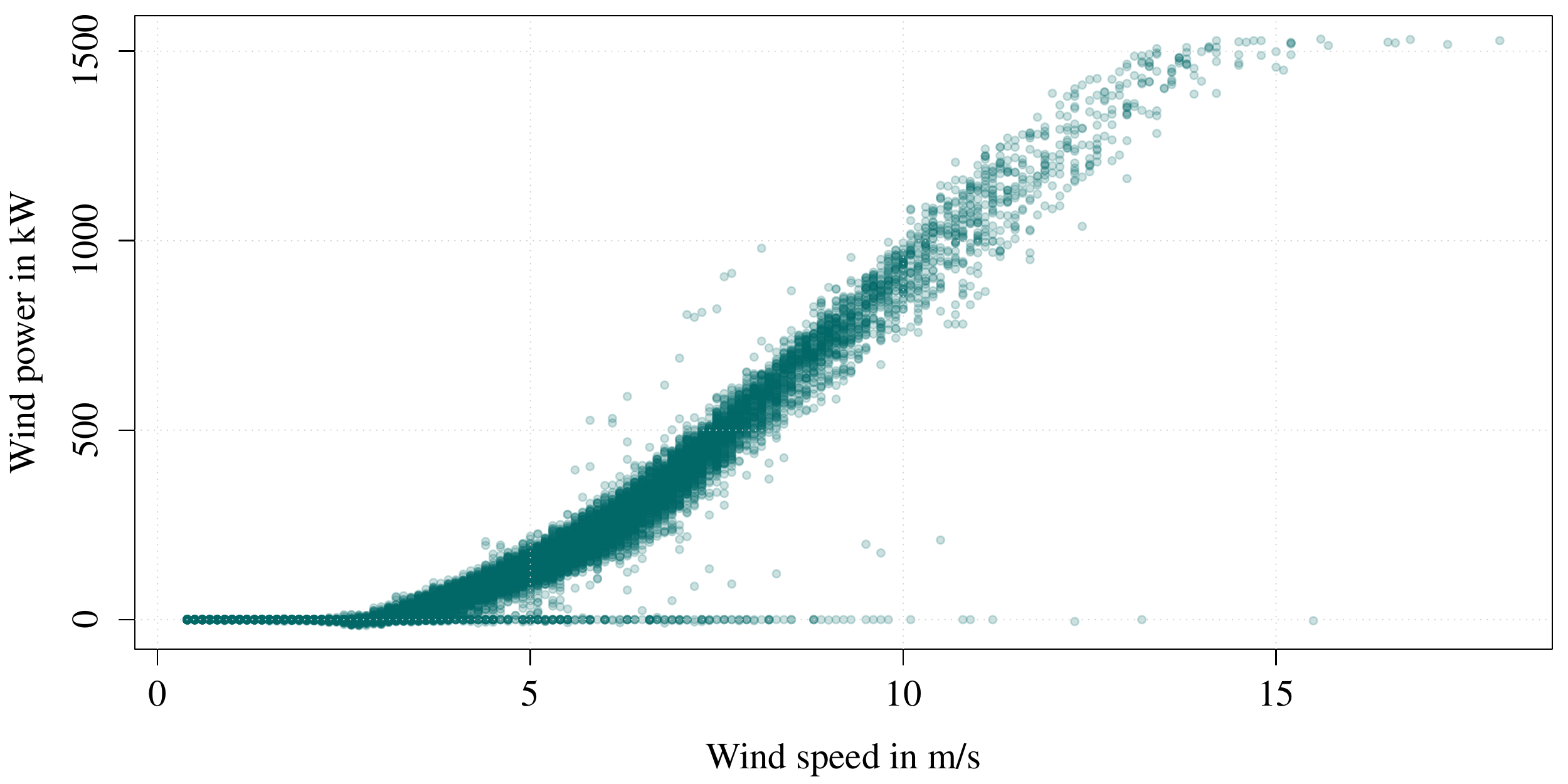}
   \caption{Turbine A.}
\end{subfigure}
\begin{subfigure}[b]{0.49\textwidth}
 \includegraphics[width=1\textwidth]{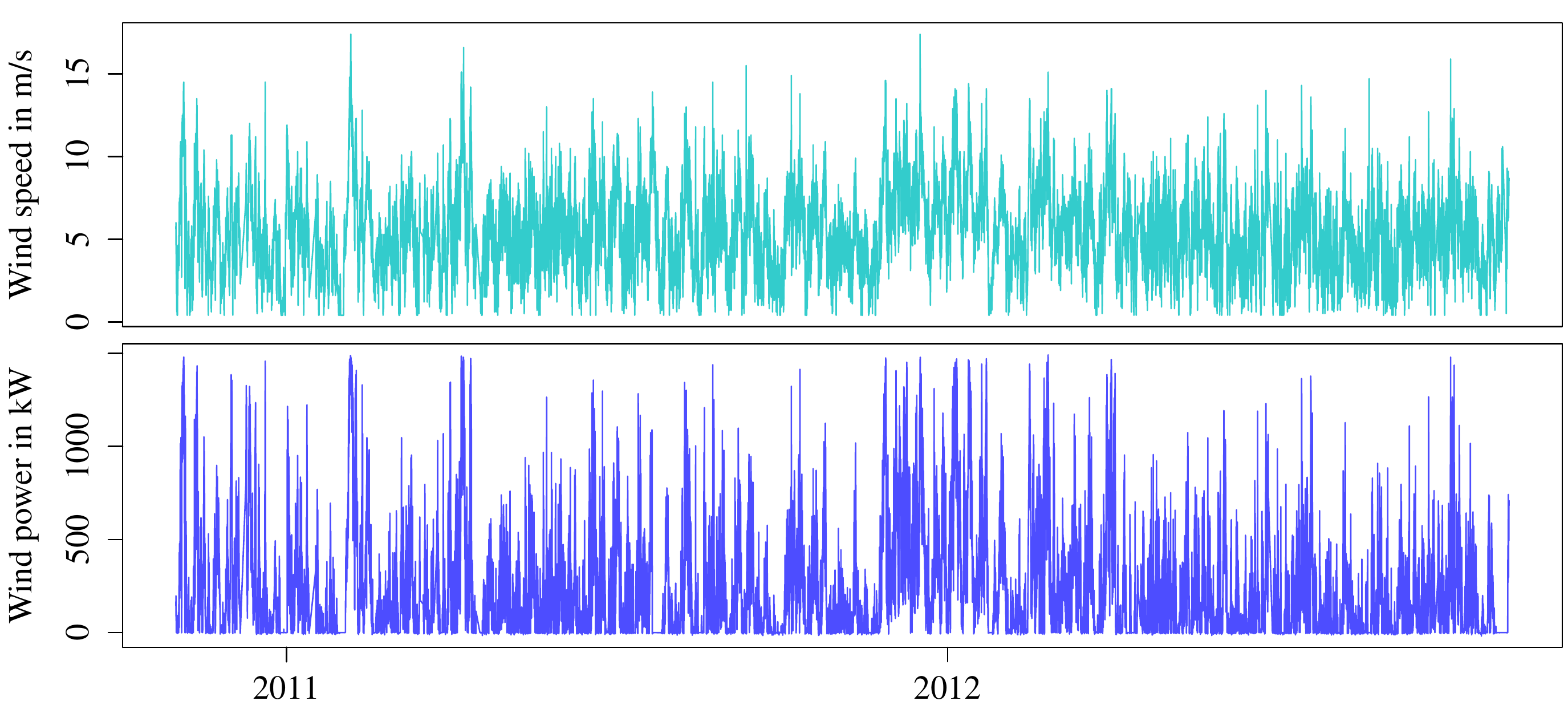}
 \includegraphics[width=.49\textwidth]{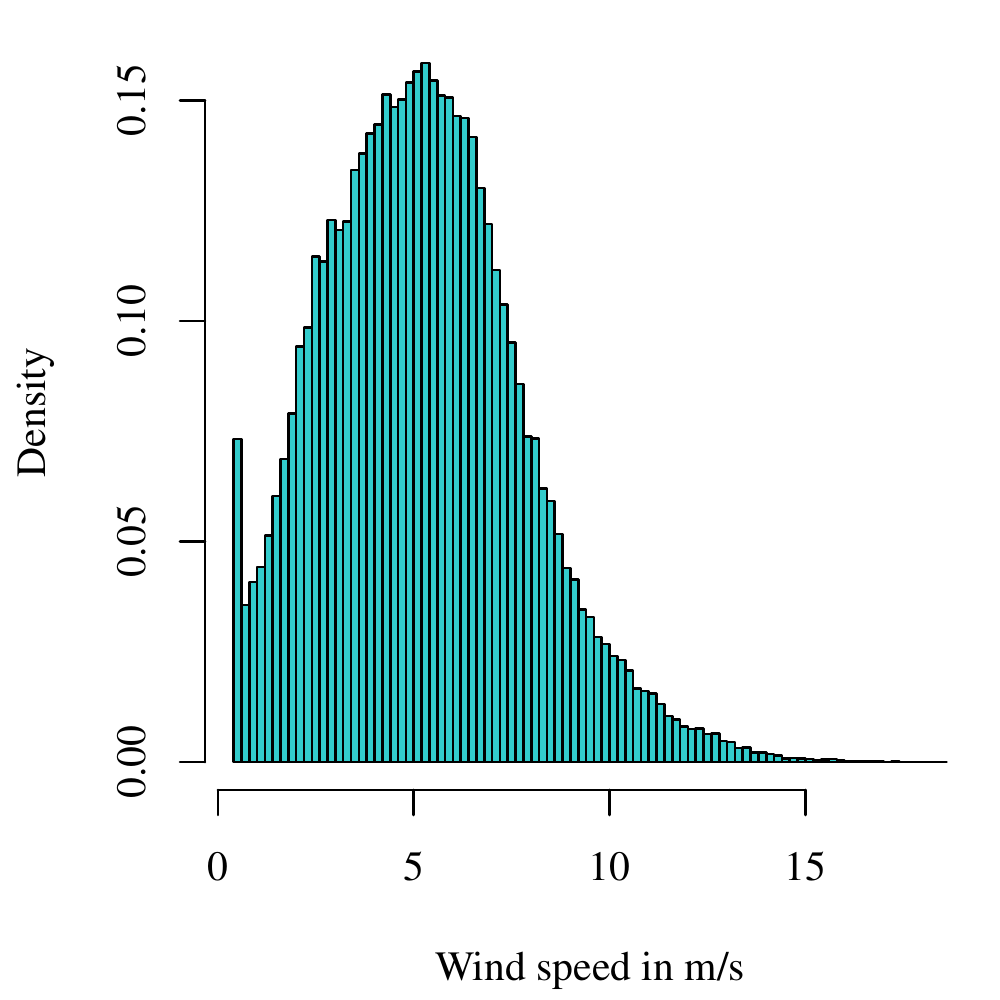}
 \includegraphics[width=.49\textwidth]{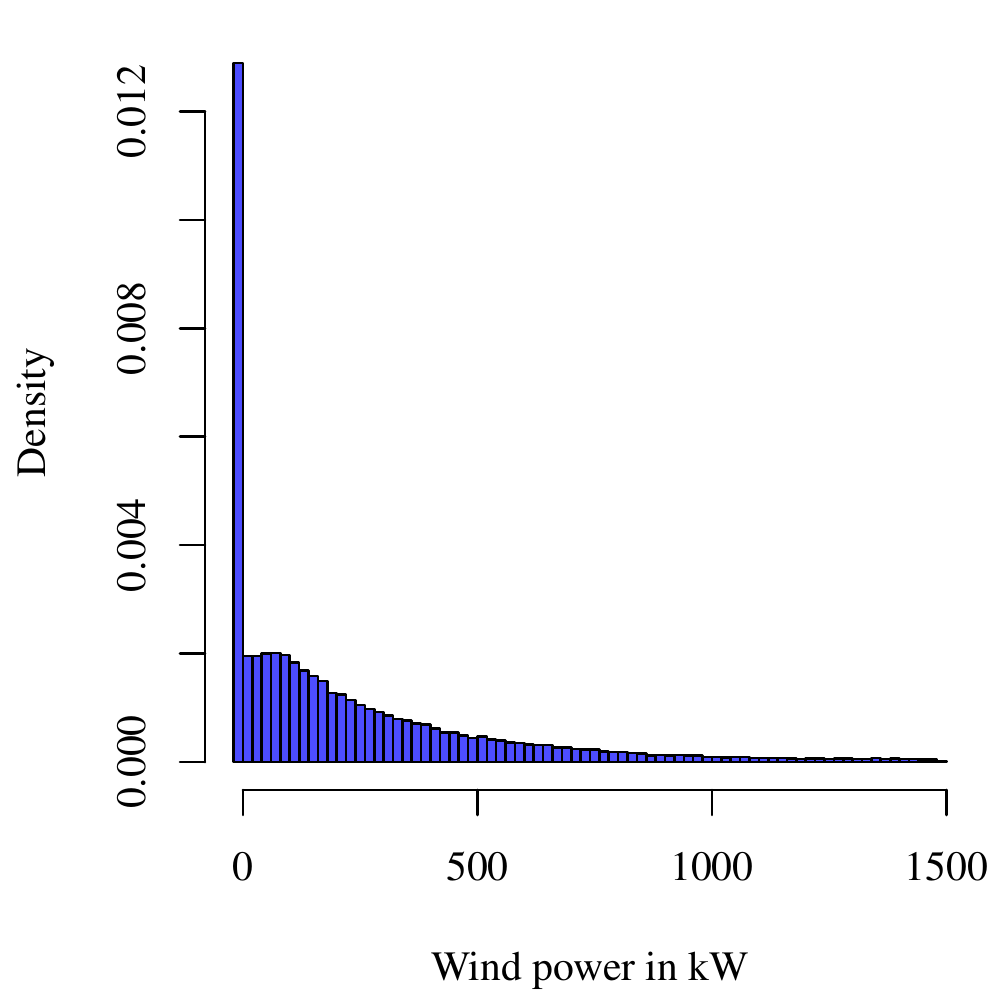}
 \includegraphics[width=1\textwidth]{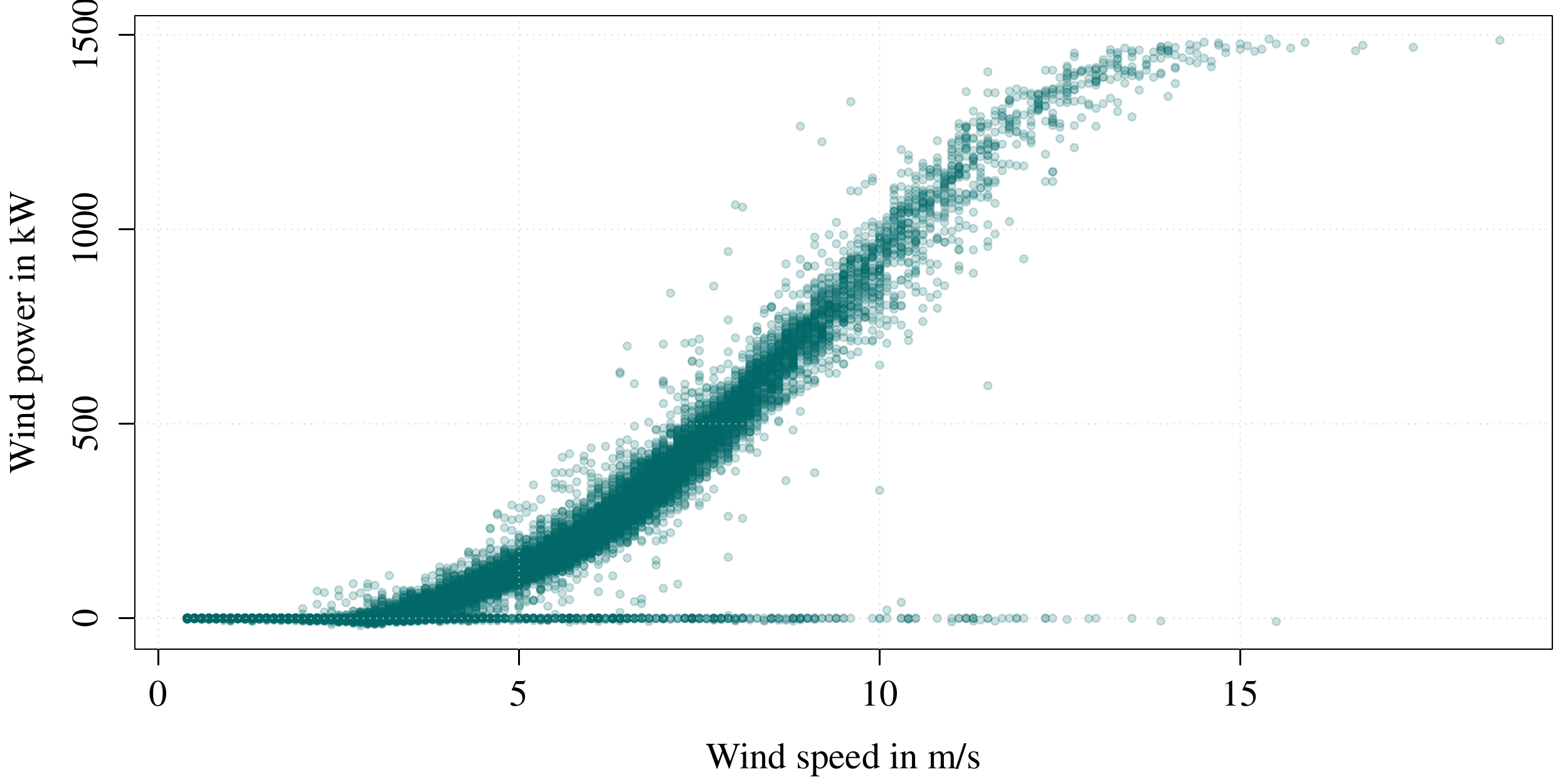}
   \caption{Turbine B.}
\end{subfigure}
 \caption{Time series, histograms of wind speed and power and corresponding empirical power curves of Turbines A and B.}
 \label{fig_means}
\end{figure}

Table \ref{table:Descriptives} shows descriptive statistics for two
of the turbines in the data set. Note that wind power observations
may very well be slightly below zero: If wind speeds are below
cut-in speed (i.e. there is no hub rotation and thus, no power
production), the turbine consumes power for system operation and
aviation lights. Also, nacelle and rotor pitch adjustments require
appreciable amounts of electricity and are mostly performed during
times at which the turbines do not produce power themselves. Thus,
some power observations are below zero. Considering the entire data
range (from -19 to 1542), the part of values below zero is only
about $1 \%$ of the range. Thus, we do not consider turbine power
consumption to be of much importance. The histograms in Figure
\ref{fig_means} support that determination. This also holds true for
the very few observations at which the theoretical power maximum of
1500 kW is exceeded. Hence, we stay with the theoretical range
assumption of $[0;1500]$ kW.\\
The data set possesses a minor number of missing values due to
engine error, maintenance shutdown or ice error. About $3 \%$ of the
data are missing, but the gaps are small: The maximum run length of
missing values is 586, which is about $0.007 \%$ of the entire data
set. Therefore, we easily fill the gaps by simple linear interpolation.\\
A scatter plot (empirical power curve) and time series plots of wind
speed and wind power are given in Figure \ref{fig_means}. It shows
that wind speed and wind power follow a similar structure and are
closely interdependent.

\begin{table}[ht]
\centering
\begin{tabular}{rrrrrrrrrr}
  \toprule
 Statistic & Min  & Median & Max & Mean & SD \\
  \midrule
 Speed A & 0.4  & 5.2 &  18.0 & 5.1 & 2.4   \\
   Power A & -19.0  & 150.0  & 1532.0 & 217.2 & 272.0   \\
 \midrule
   Speed B  & 0.4  & 5.5 & 18.6 & 5.3 & 2.5   \\
   Power B  & -19.0  & 155.0  & 1493.0 & 230.5 & 291.1   \\
 \bottomrule
\end{tabular}
 \caption{Descriptive statistics of Turbines A and B. Wind speed denoted in m/s, wind power in kW.}
   \label{table:Descriptives}
\end{table}

\begin{figure}[htb!]
  \centering
  \begin{subfigure}[b]{0.49\textwidth}
  \includegraphics[width=1\textwidth]{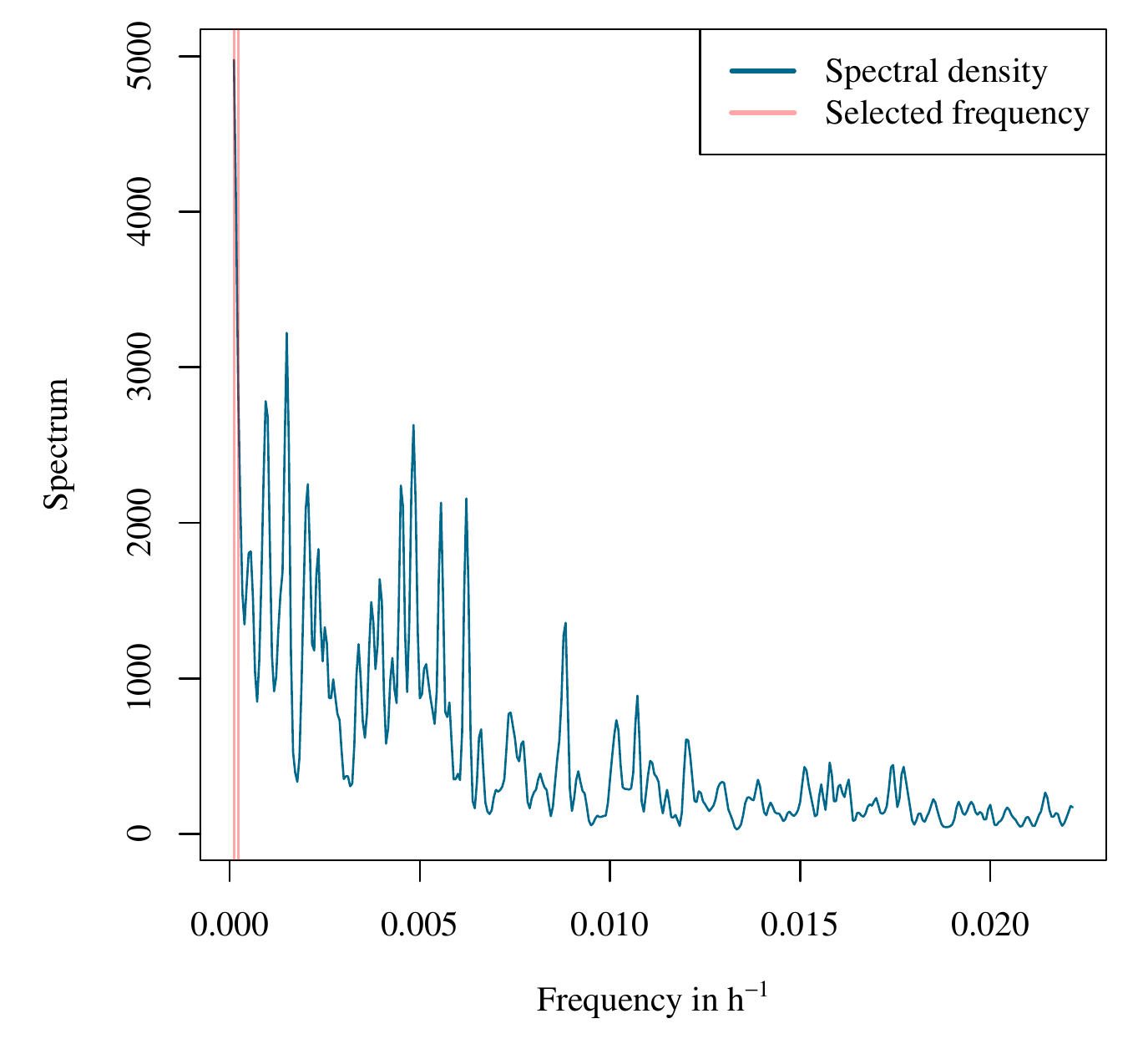} \\
  \includegraphics[width=1\textwidth]{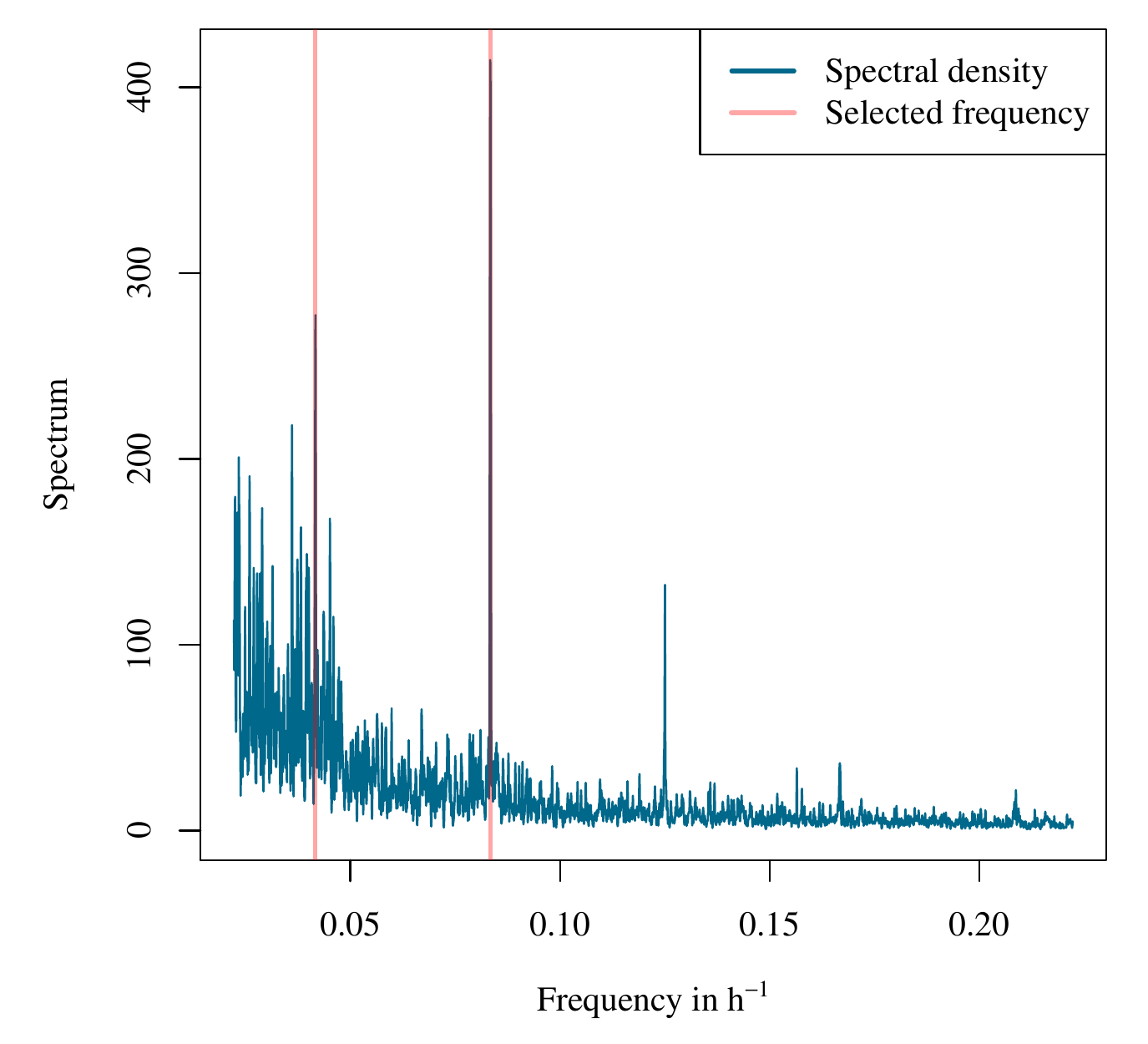}
    \caption{Smoothed Periodogram of wind speed for Turbine A.}
   \label{graph:SpecWS}
  \end{subfigure}
  \begin{subfigure}[b]{0.49\textwidth}
   \includegraphics[width=1\textwidth]{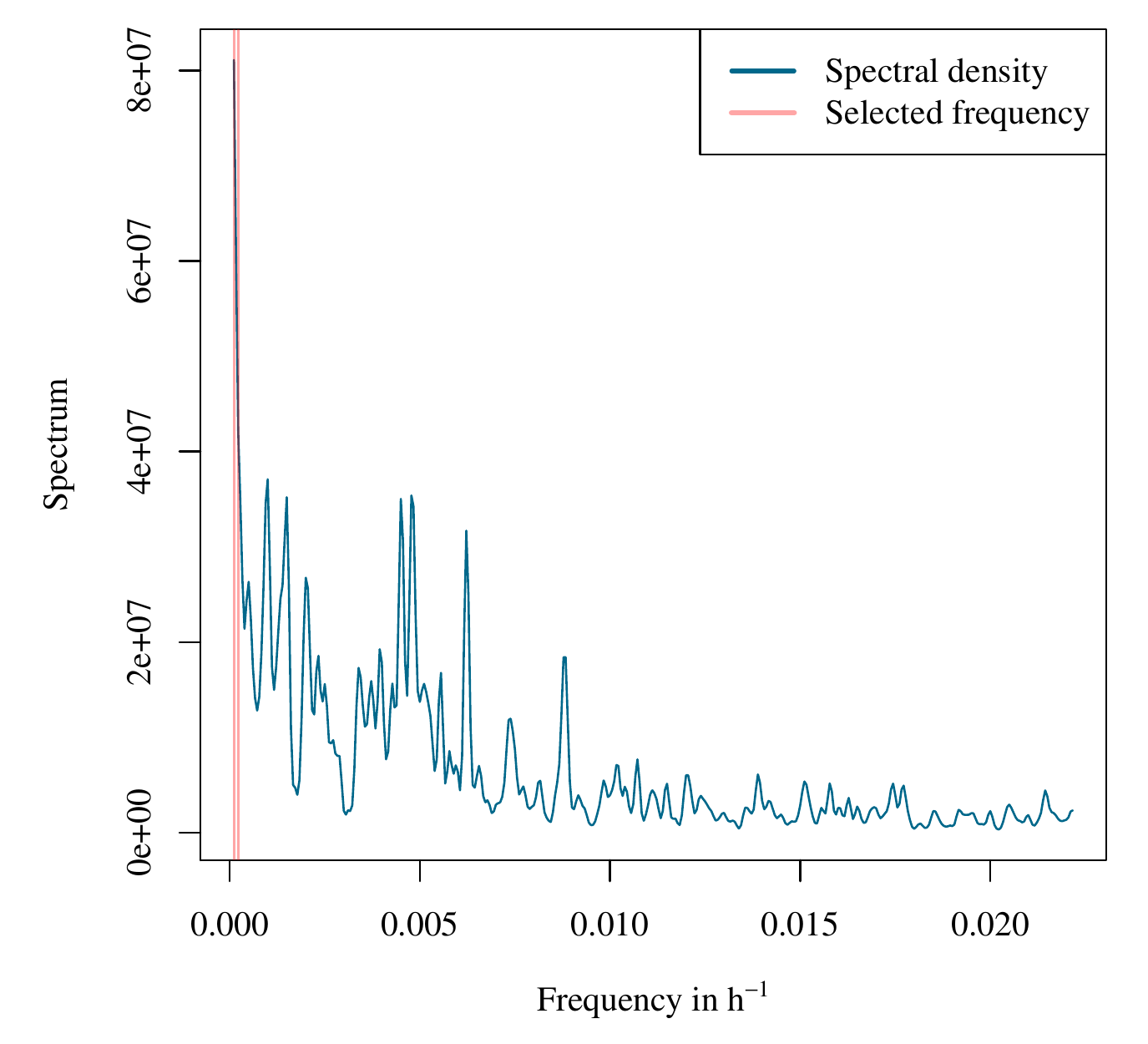} \\
   \includegraphics[width=1\textwidth]{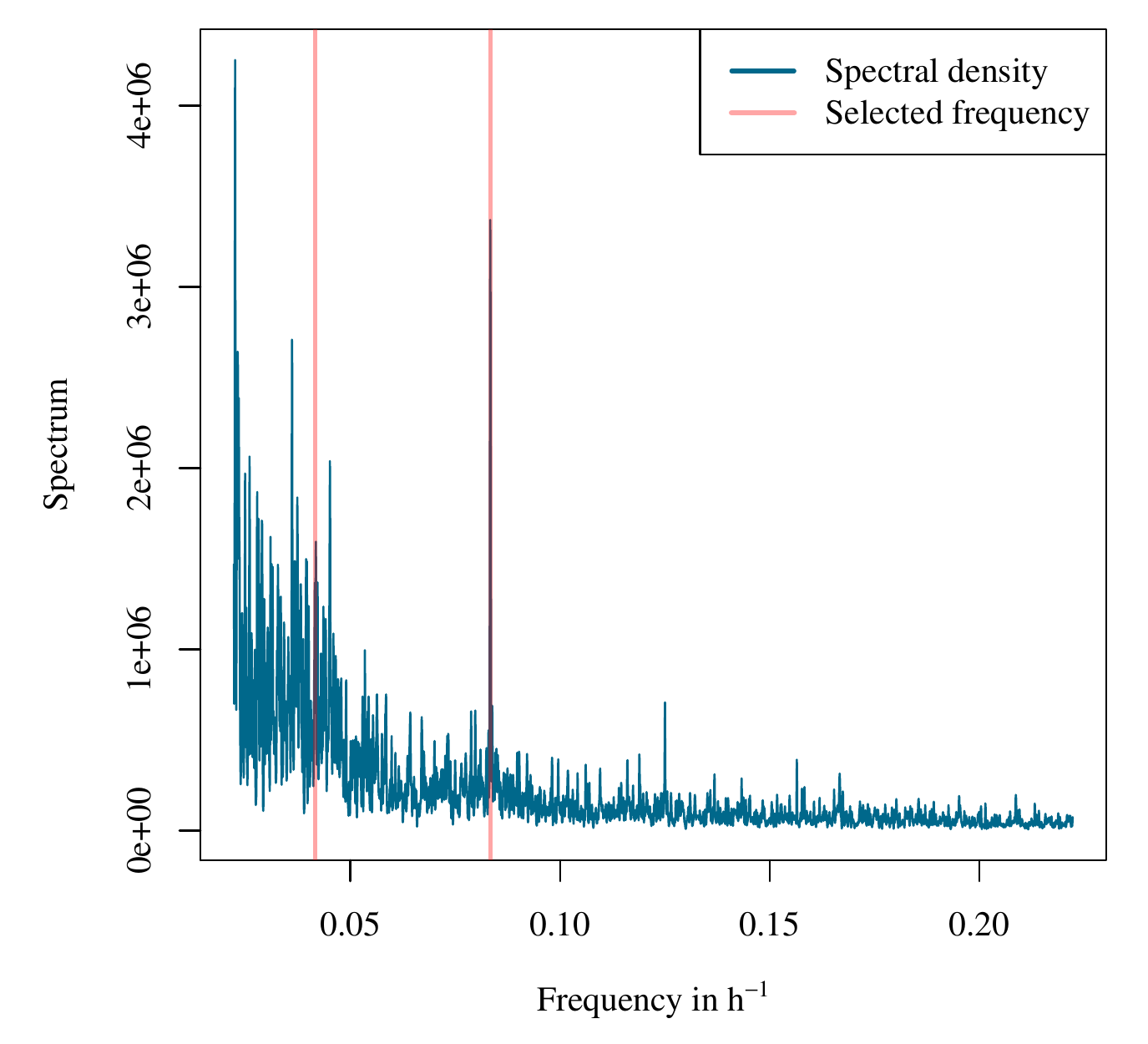}
    \caption{Smoothed Periodogram of wind power for Turbine A.}
   \label{graph:SpecWP}
 \end{subfigure}
  \caption{Estimated spectral density of wind speed (left panels) and wind power (right panels).}
  \label{graph:Spec}
\end{figure}

Besides the high persistence of wind speed and wind power which is directly related to the high-frequency data set, it is necessary to discuss another important characteristic of wind power and wind speed. The wind speed data provides a strong periodic behavior, as \cite{zhu2014space} and \cite{ambach2015periodic} point out. The wind power data set also provides these characteristics. A diurnal periodicity as considered for the WPPT and \mbox{GWPPT} is observable for our data set, but the annual period is not completely straight forward. Hence, we calculate the sample periodogram, which is shown in Figure \ref{graph:Spec}.\\
Figure \ref{graph:Spec} shows the estimated spectral density for the
wind power and wind speed of turbine A. The red lines in Figure
\ref{graph:Spec} show annual and half-annual frequencies in the
upper panels and daily and half-daily periods in the lower panels.
The diurnal periodicity is not so prevalent within the wind power
series shown in the right-hand panels, but there are several
important frequencies nearby the daily period. After all, periodic
B-spline functions will help to model all multiples of a diurnal and
annual periodicity.

\section{Model}\label{section:Model}

Let $d=8$ denote the number of turbines in the wind park. Thus, the
set of turbines is $\DD = \{1, \ldots, d\}$. The $d$-dimensional
time series of wind speed is $(\bsW_t)_{t\in \Z}$ with $\bsW_t =
(W_{1,t}, \ldots, W_{d,t})'$
and the wind power is $(\bsP_t)_{t\in \Z}$ with $\bsP_t = (P_{1,t}, \ldots, P_{d,t})'$.\\
We split the model description into two parts: First, we present the
multivariate time varying threshold VARMA model for the wind speed.
Its heteroscedastic variance structure is modeled by a TGARCH type
process. Afterward, we present the wind power model which includes
the wind speed dependence and considers the errors of wind power
themselves to follow a power-TGARCH process.

\subsection{The Wind Speed Component}\label{subsection:Wind Speed}

For the wind speed $\bsW_t$ we consider the multivariate time
varying threshold-VARMA model

\begin{align}
 W_{i, t} = \phi_{i, 0}(t)  &+ \sum_{j \in \DD} \sum_{k \in I^\phi_{i, j}} \sum_{c \in C^{\phi}_{i, j, k}}
 \phi_{i, j, k, c}(t) \max\{W_{j, t - k}, c\} \nonumber \\
 &+ \sum_{j \in \DD} \sum_{k \in I^\theta_{i, j}} \theta_{i, j, k}(t) \eps_{j, t - k} + \eps_{i, t},
 \label{eq_wind_ar_model}
\end{align}

where $i \in \DD$, $\phi_{i, j, k, c}$ resp. $\theta_{i, j, k}$ represent the time varying autoregressive and moving average coefficients and $\eps_{i, t}$ is the error term. The index sets $I^{\phi}_{i, j}$ and $I^{\theta}_{i, j}$ contain the corresponding relevant AR- and MA-lags and the threshold set $C^{\phi}_{i, j, k}$ contains all considered thresholds in the autoregressive part. The simple choice $C^{\phi}_{i, j, k} = \{-\infty\}$ would turn the model into a standard time varying VARMA process. The thresholds describe the AR-dependence of wind speed by a piecewise linear function with breaks at the corresponding thresholds. Just as each smooth function, it can be approximated well by piecewise linear functions, which provides a flexible and efficient way to capture the non-linear dependence in the data.\\
We assume the error process $(\eps_{i, t})_{t\in \Z}$ to be
conditionally heteroscedastic. Therefore, we consider $\eps_{i, t} =
\sigma_{i, t} Z_{i, t}$, where $(Z_{i, t})_{t \in \Z}$ is i.i.d.
with $\E(Z_{i, t}) = 0$ and $\var(Z_{i, t}) = 1$. In detail, we
assume that $\eps_{i, t}$ follows a time varying TGARCH process,
such that

\begin{align}
\sigma_{i, t} = \alpha_{i, 0}(t) & + \sum_{j \in \DD} \sum_{k \in I^{\alpha}_{i, j}}
 \alpha^+_{i, j, k}(t) \eps^{+}_{j, t - k} + \alpha^-_{i, j, k}(t) \eps^{-}_{j, t - k} \nonumber \\
&+ \sum_{j \in \DD} \sum_{k \in I^{\beta}_{i, j}} \beta_{i, j, k}(t) \sigma_{j, t - k}
 \label{eq_wind_arch_model}
\end{align}

with index sets $I^{\alpha}_{i, j}$ and $I^{\beta}_{i, j}$, $\eps^{+}_{j, t - k} = \max\{ \eps_{j, t - k}, 0\}$, $\eps^{-}_{j, t - k} = \max\{ -\eps_{j, t - k}, 0\}$  and time varying coefficients $\alpha_{i, 0}(t) > 0$, $\alpha^+_{i, j, k}(t)\geq 0$, $\alpha^-_{i, j, k}(t)\geq 0$, $\beta_{i, j, k}(t)\geq 0$. The index sets for the considered lags are given in Table \ref{tab_lags}.\\
For most of the coefficients, we allow dependence of up to one hour (6 lags) only to keep the specification manageable. However, for the coefficients that describe the wind dependence on its own past, we allow for more parameters. Here, we also include the lags $140, \ldots, 150$ to cover the impact from the previous day (which corresponds to $6 \times 24 = 144$ lags).\\
For the thresholds $C^{\phi}_{i, j, k}$, we use a parsimonious lag
specification: We allow non-linear impacts for the first two lags,
only. The elements of $C^{\phi}_{i, j, k}$ contain the $10\%$
percentiles of the process in the mean equation. Thus, $C_{i, j,
k}^{\phi}$ contains the $10\%$ percentiles in the cases $k = 1$ or
$k = 2$. All elements that do not satisfy this restriction are set
to $C^{\phi}_{i, j, k} = \{-\infty\}$. The non-linear impact of the
threshold model specification acts via piecewise linear functions to
cover possibly present turbulent flow and wake effects. The effect
of this model component is explained in detail in the following
subsection.\\
To keep the parameter space reasonable, we keep most of the
coefficients constant and allow only a few important ones to vary
over time. For $I_{i, j}^{\phi}$ as well as $I_{i, i}^{\theta}$, we
consider the coefficients for lags 1 and 2 to be time varying.
\cite{ziel2015efficient} proceed similarly for modeling the wind and
solar power net feed-in.

\begin{table}[tbh]
\centering
\begin{tabular}{ll}
\toprule
Index sets & Contained lags \\
\midrule
$I_{i, i}^{\phi}$, $I_{i, i}^{\alpha}$ & 1, \ldots, 40 \text{and} 140, \ldots, 150 \\
$I_{i, j}^{\phi}$, $I_{i, i}^{\theta}$, $I_{i, j}^{\theta}$, $I_{i, j}^{\alpha}$,$I_{i, i}^{\beta}$, $I_{i, j}^{\beta}$ & 1, \ldots, 6 \\
\bottomrule
\end{tabular}
\caption{Considered lags of the index sets, where $i, j \in \DD$ with $j \neq i$.}
\label{tab_lags}
\end{table}

\subsection{The Wind Power Model}\label{subsection:Wind Power}

For the wind power process we assume a model that is given by

\begin{align}
 P_{i, t} =& \varphi_{i, 0}(t)
 + \sum_{j \in \DD} \sum_{k \in I^\varphi_{i, j}} \sum_{c \in C^{\varphi}_{i, j, k} } \varphi_{i, j, k, c}(t) \max\{ P_{j, t - k}, c\} \nonumber \\
 &+ \sum_{j \in \DD} \sum_{k \in I^\psi_{i, j}} \sum_{c \in C^{\psi}_{i, j, k} } \psi_{i, j, k, c}(t) \max\{ W_{j, t - k}, c\}
 \nonumber \\
 &+ \sum_{j \in \DD} \sum_{k \in I^\vartheta_{i, j}} \vartheta_{i, j, k}(t) \epsilon_{j, t - k}   \nonumber \\
 &+ \sum_{j \in \DD} \sum_{k\in I^\varpi_{i, j}} \varpi_{i, j, k}(t) \eps_{j, t - k} +  \epsilon_{i, t}.
 \label{eq_power_ar_model}
\end{align}

\begin{figure}[hbt!]
\centering
\begin{subfigure}[b]{0.49\textwidth}
 \includegraphics[width=1\textwidth]{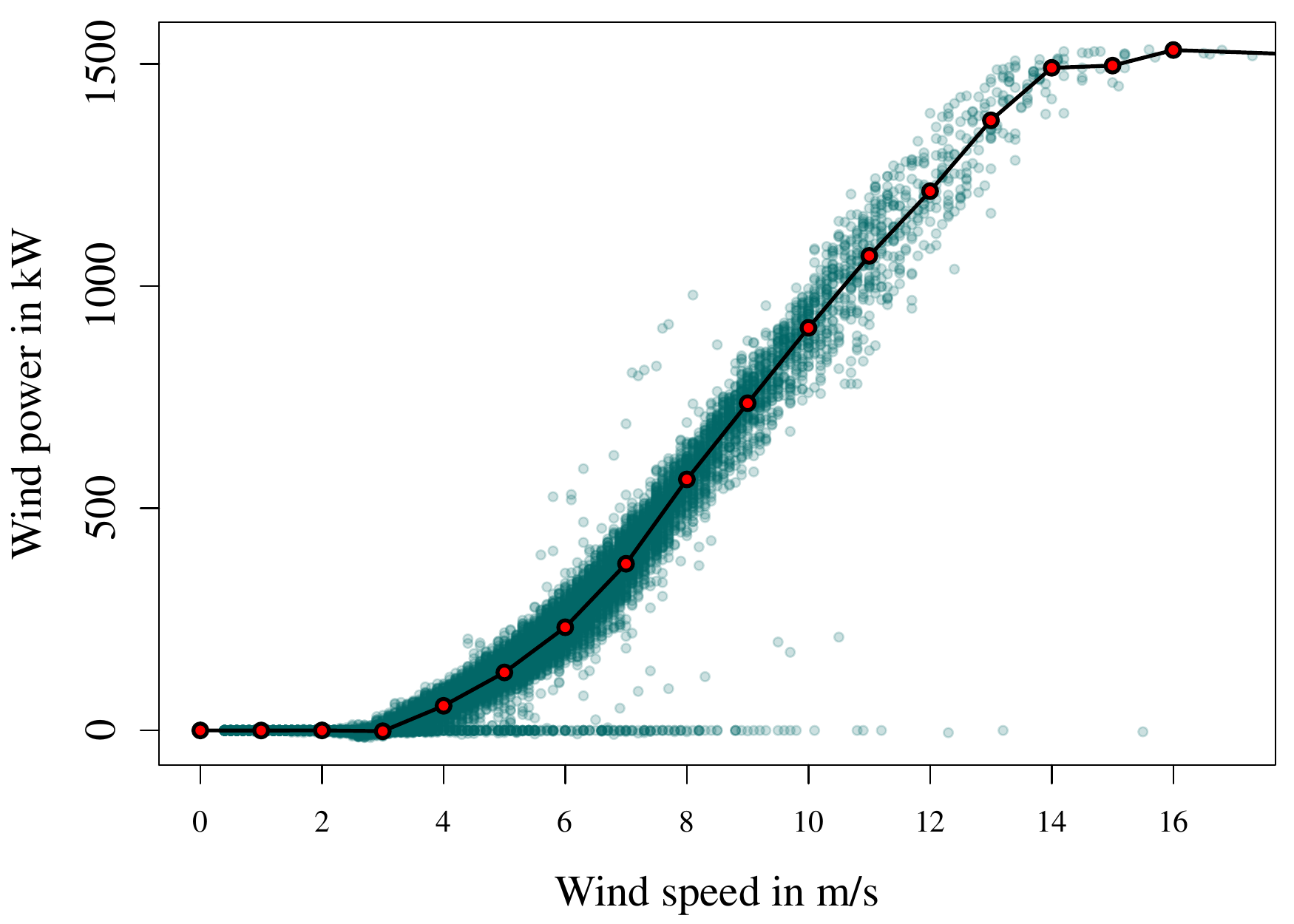}
   \caption{Turbine A.}
  \label{fig_threshB}
\end{subfigure}
\begin{subfigure}[b]{0.49\textwidth}
 \includegraphics[width=1\textwidth]{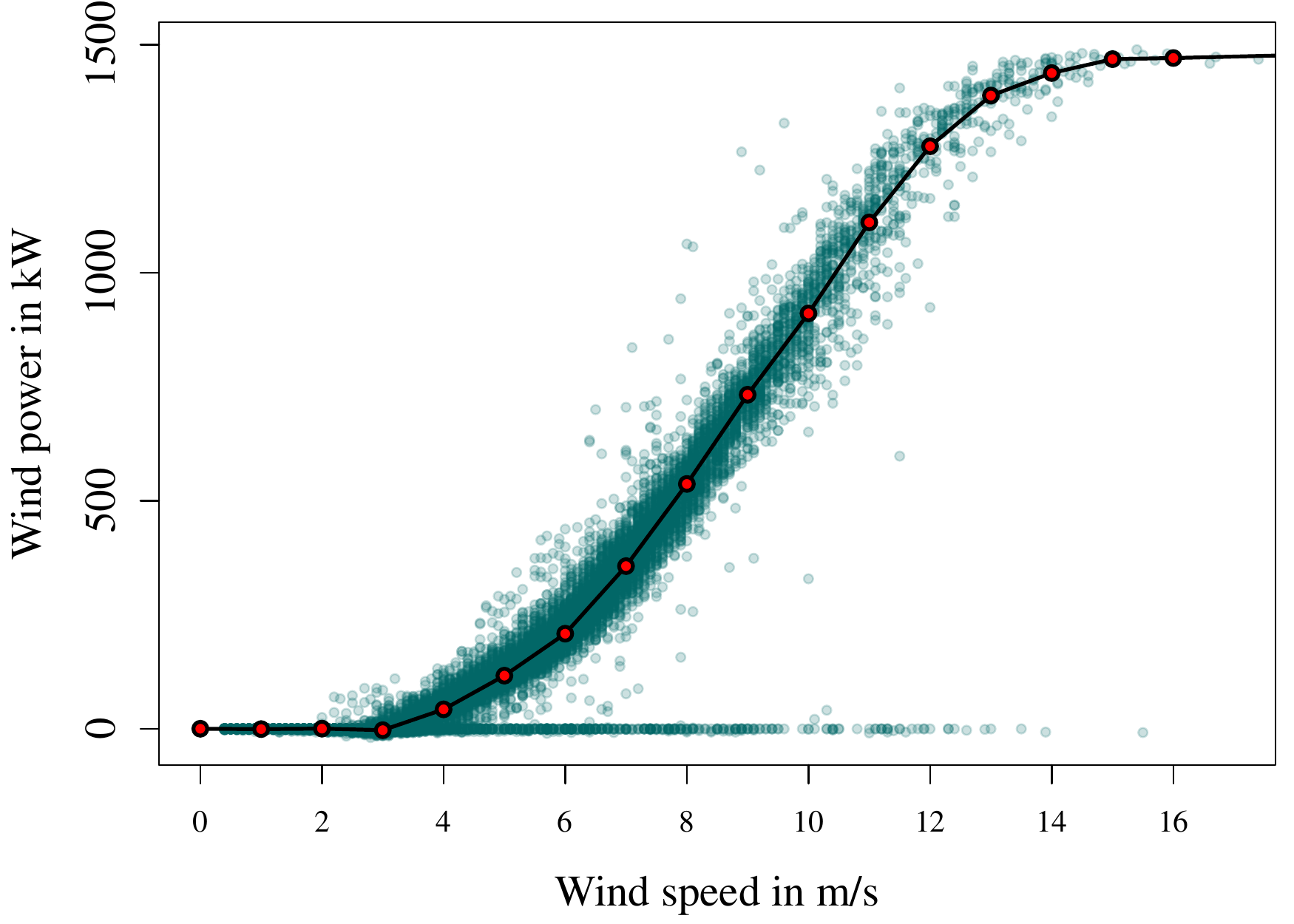}
   \caption{Turbine B.}
  \label{fig_threshB}
\end{subfigure}
 \caption{Fitted results of the illustrative example of model \eqref{eq_threshold_example}.}
 \label{fig_thresh}
\end{figure}

All observed turbines are located in close proximity to each other,
and are influenced by the same air pressure and weather conditions.
Dependent on the angle of movement of a particular pressure area
(and thus, wind conditions) at any one time, these conditions may
hit one set of turbines sooner than others. The spatial dispersion
of the turbines' power production can therefore be accounted for by
a time-lag structure. Thus, we assume that the power $P_i$ of turbine $i$ can depend on its own past as well as on the past of the power of the other turbines by the $\varphi$ parameters. The power can also depend on the current and past wind speed by means of the $\psi$ parameters. Furthermore, wind power depends on the past residuals. Note that the lag structure for the wind power model is slightly different from that of the wind speed model, as we assume a causal/temporal structure in the data. We assume that the wind speed $\bsW_t$ at time $t$ can only depend on the past wind speed $\bsW_{t - k}$ for $k \geq 1$. Similarly, the wind power $\bsP_t$ depends on the past wind power $\bsP_{t - k}$ for $k \geq 1$, but also on current and past wind speed $\bsW_{t - k}$ for $k \geq 0$.\\
Comparably to the wind speed model, we allow for non-linear wind speed and wind power effects by several thresholds. Particularly, the theoretical non-linear effect of the wind speed on the wind power is well known to be described by the third-degree polynomial:

\begin{equation}
P = \frac{1}{2}\rho C_P A W^3,
\label{eq_wpd}
\end{equation}

where $\rho$ describes the air density, $C_P$ denotes the physical
properties of the turbine (values of up to 16/27, the so-called Betz
limit), and $A$ represents the swept area. \cite{Hennessey1977} goes
into details. However, especially around the upper bound of the
maximum produced wind power, it is known that the true impact of
wind speed on wind power is different from the usual cubic
relationship and should not be modeled to be cubic. Our way of
modeling the non-linear impact by piecewise linear effects allows
for a flexible way to model the underlying non-linear impact. To
illustrate the impact of the thresholds we briefly present a simple
threshold model for the wind power dependent on the wind speed. It
is given by

\begin{align}
 P_{i, t} =& a + \sum_{c=0}^{16} b_{c} \max\{ W_{i, t}, c\} + \text{e}_{i,t}
 \label{eq_threshold_example}
\end{align}

for turbine $i$. It contains thresholds at $0,1,2, \ldots, 16$ for
modeling the non-linear relationship by piecewise linear functions.
In Figure \ref{fig_thresh}, the fitted values of model
\eqref{eq_threshold_example} are given for Turbines A and B of the
investigated wind park. It can be seen that the piecewise linear
approach is able to cover the non-linear relationship quite well.
There are distinct bents at the threshold points. In fact, the
fitted curve is a linear spline. Of course, a higher number of
thresholds will increase the model fit. However, if the number of
thresholds is too large, it might lead to overfitting. Still, this
problem is somehow limited due to our shrinkage estimation
procedure, so that a large number of parameters in the problem space
does not necessarily imply a lot of estimations in the solution.
Instead, the algorithm will automatically select the most plausible
piecewise linear function that approximates the non-linear impact
well, as we choose the thresholds in model \eqref{eq_power_ar_model} to be data driven.\\
Similarly as for the wind speed process, we assume a GARCH-type
process for the wind power error, so $\epsilon_{i, t} =
\varsigma_{i, t} U_{i, t}$ with $U_{i, t}$ i.i.d., $\E(U_{i, t}) =
0$ and $\E(U_{i, t}^2) = 1$. This slightly differs from the TGARCH
process for the wind speed: We assume that the third-degree
relationship in equation \eqref{eq_wpd} hands down to the residual
volatility. Thus, instead of considering a recursion on
$\varsigma_{i, t}$, we consider a recursion on the cubed root of the
volatility $\varsigma_{i, t}^{\frac{1}{3}}$. Consequently, we assume
that $\epsilon_{i, t}$ follows a time varying power-TGARCH process:

\begin{align}
\varsigma_{i, t}^{\frac{1}{3}} = &\eta_{i, 0}(t) + \sum_{j \in \DD} \sum_{k \in I^{\eta}_{i, j}} \eta^+_{i, j, k}(t) |\epsilon^{+}_{j, t - k}|^{\frac{1}{3}} + \eta^-_{i, j, k}(t) |\epsilon^{-}_{j, t - k}|^{\frac{1}{3}} \nonumber \\
&+ \sum_{j \in \DD} \sum_{k \in I^{\zeta}_{i, j}} \zeta_{i, j, k}(t)
\varsigma_{j, t - k}^{\frac{1}{3}}
  + \sum_{j \in \DD} \sum_{k \in I^{\upsilon}_{i, j}} \upsilon^+_{i, j, k}(t) |\eps^{+}_{j, t - k}|^{\frac{1}{3}}  \nonumber \\
  &+ \upsilon^-_{i, j, k}(t) |\eps^{-}_{j, t - k}|^{\frac{1}{3}} + \sum_{j \in \DD} \sum_{k \in I^{\varrho}_{i, j}} \varrho_{i, j, k}(t) \sigma_{j, t - k}^{\frac{1}{3}},
 \label{eq_power_arch_model}
\end{align}

$\epsilon^{+}_{j, t - k} = \max\{ \epsilon_{j, t - k}, 0 \}$,
$\epsilon^{-}_{j, t - k} = \max\{ -\epsilon_{j, t - k}, 0 \}$.
Finally, the index sets for the considered lags on the mean part of
the model in equation \eqref{eq_power_ar_model} and the variance
part in equation \eqref{eq_power_arch_model} are given in Table
\ref{tab_lags_power}. The corresponding parameters for the index
sets $I_{i, j}^{\alpha}$, $I_{i, j}^{\beta}$, $I_{i, j}^{\varphi}$,
$I_{i, j}^{\vartheta}$, $I_{i, j}^{\eta}$, $I_{i, j}^{\upsilon}$,
$I_{i, j}^{\zeta}$ and $I_{i, j}^{\varrho}$ are considered to be
time varying on lags $1$ and $2$, those for the sets $I_{i,
j}^{\psi}$ and $I_{i, j}^{\varpi}$ are time varying on lags $0, 1,
2$ and the corresponding regressors are modeled by periodic
B-splines, as discussed subsequently.

\begin{table}[tbh]
\centering
\begin{tabular}{p{3.2cm}p{4cm}}
\toprule
Index sets & Contained lags \\
\midrule
$I_{i, i}^{\varphi}$, $I_{i, i}^{\eta}$, $I_{i, i}^{\upsilon}$ & 1, \ldots, 40 \text{and} 140, \ldots, 150 \\
$I_{i, i}^{\psi}$  & 0, \ldots, 40 \text{and} 140, \ldots, 150 \\
$I_{i, i}^{\vartheta}$, $I_{i, j}^{\varphi}$, $I_{i, j}^{\vartheta}$, $I_{i, j}^{\eta}$, $I_{i, j}^{\upsilon}$, $I_{i, i}^{\zeta}$, $I_{i, j}^{\zeta}$, $I_{i, i}^{\varrho}$, $I_{i, j}^{\varrho}$ & 1, \ldots, 6 \\
$I_{i, i}^{\varpi}$, $I_{i, j}^{\psi}$, $I_{i, j}^{\varpi}$ & 0, \ldots, 6 \\
\bottomrule
\end{tabular}
\caption{Considered lags of the index sets, where $i, j \in \DD$ with $j \neq i$.}
\label{tab_lags_power}
\end{table}

\subsection{Time Varying Coefficients}

\begin{figure}[hbt!]
\centering
\begin{subfigure}[b]{0.49\textwidth}
 \includegraphics[width=1\textwidth, height=.8\textwidth]{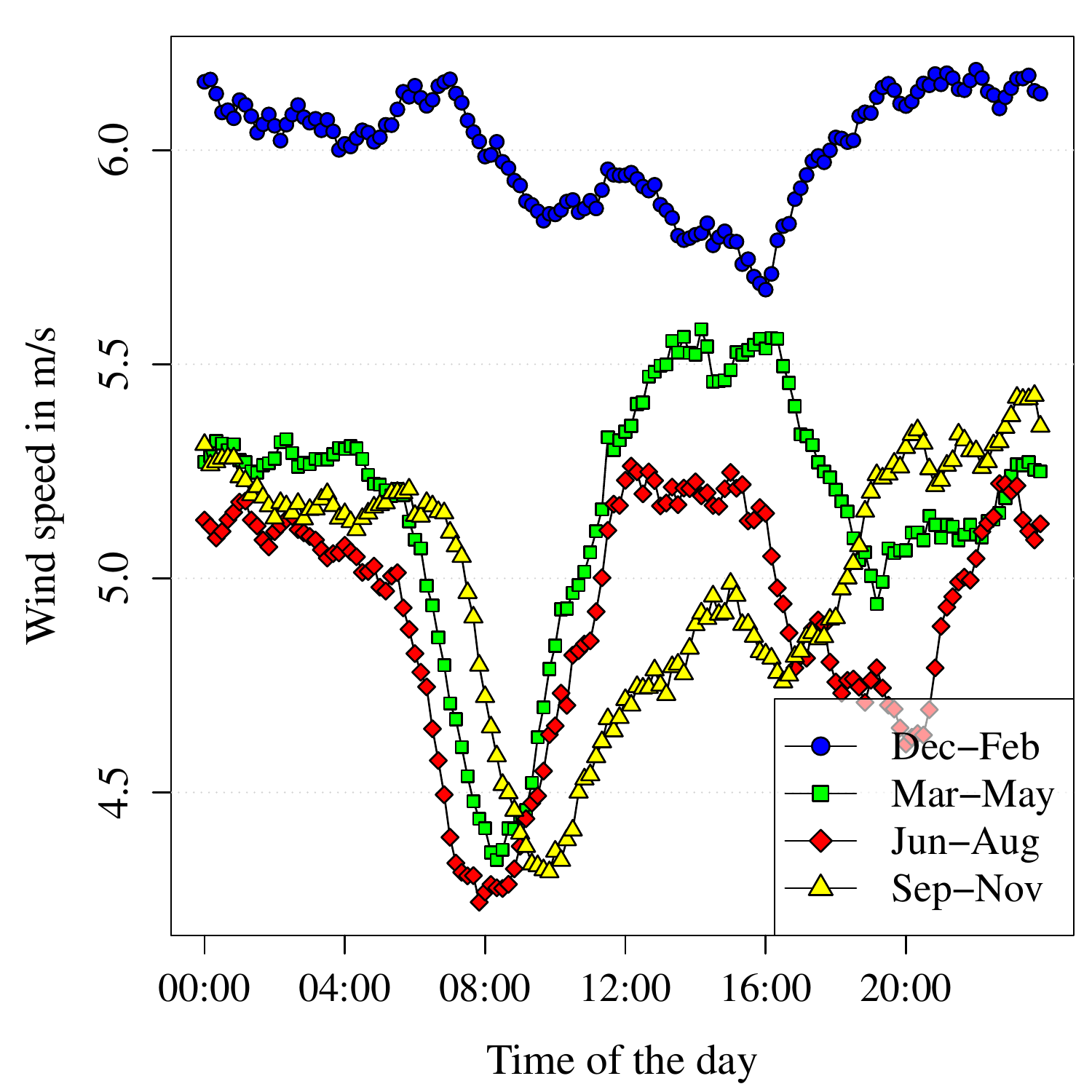}
   \caption{Mean wind speed.}
  \label{fig_m1}
\end{subfigure}
\begin{subfigure}[b]{0.49\textwidth}
 \includegraphics[width=1\textwidth, height=.8\textwidth]{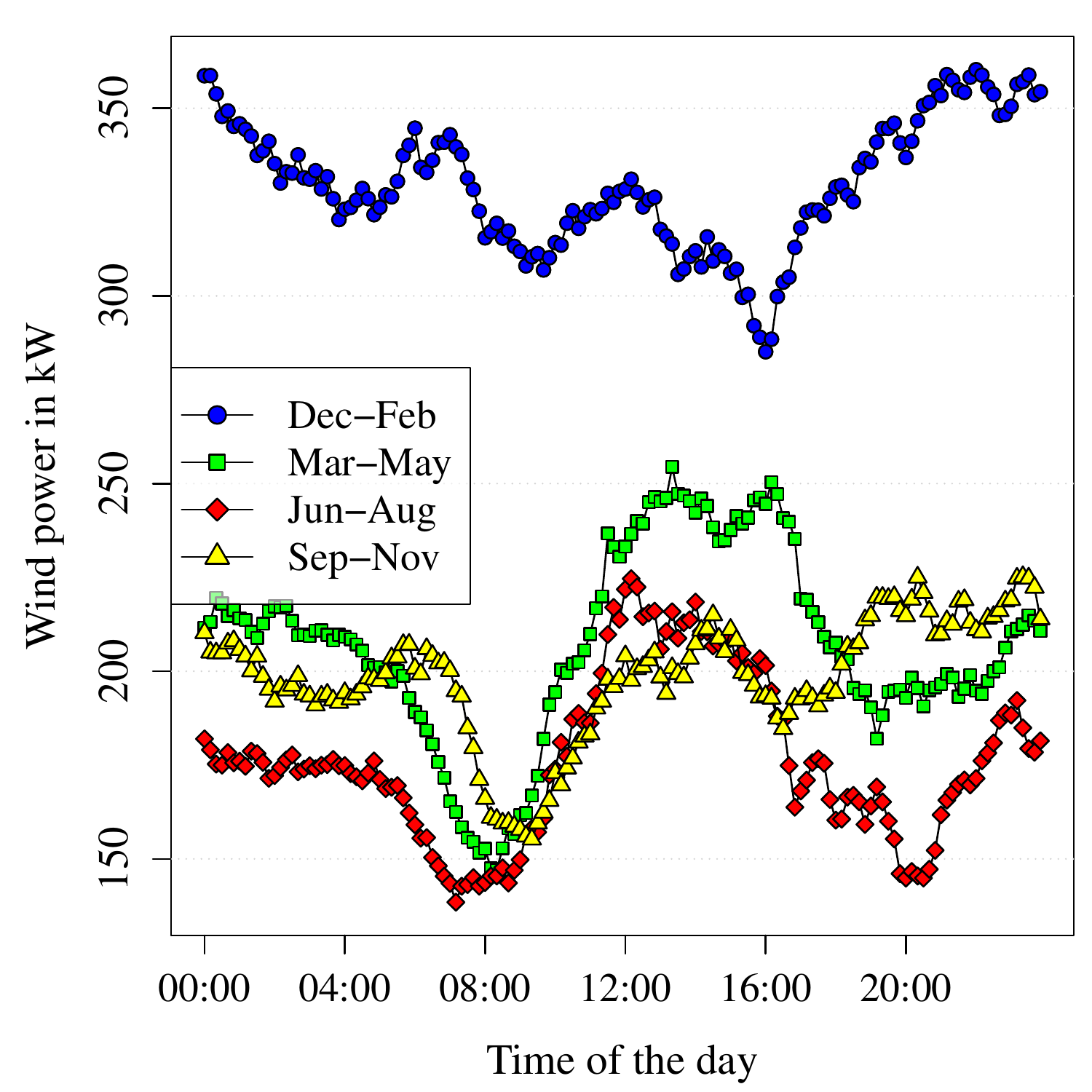}
   \caption{Mean wind power.}
  \label{fig_m2}
\end{subfigure} \caption{Daily mean wind speed and wind power of the $8$ turbines for the four seasons in a year.}
 \label{fig_means2}
\end{figure}

We assume an identical structure of the time varying coefficients in
the wind speed and in the wind power model, as both exhibit similar
seasonal effects. In general, the time varying coefficients can be
modeled by periodic functions like Fourier approximations or other
periodic basis functions, such as periodic B-splines or periodic
wavelets. We opt for the flexible cubic B-spline approach. Let $\xi$
be a time varying coefficient. Then

\begin{equation}
  \xi(t) = \sum_{l = 1}^{N_\xi} \xi_l B^{\xi}_l(t),
  \label{eq_tim-var-coef_basis}
\end{equation}

so that $\xi$ is given by a sum of $N_\xi$ basis functions $B^{\xi}_l(t)$, weighted by $\xi_l$.\footnote{Details on the construction of the B-splines basis functions are discussed in the appendix.}\\
In the literature on wind power modeling, Fourier approximations are
used frequently, see, e.g., \cite{Giebel2011}. However, the Fourier
technique is a global approach. For our purpose, a local approach is
preferable, as it is more flexible with respect to possible changes
in the time-dependent structure itself. We design our basis function
so that it can cover both the diurnal and the annual periodic
effects. Furthermore, we allow for possible interactions between
both seasonalities, so that the diurnal impact can change over the
year. This impact is visualized in Figure \ref{fig_means2}. It
displays the daily mean wind speed and wind power of all considered
wind turbines for the four seasons in a year. For both the wind
speed and the wind power, we observe that during the morning hours
around 7am to 10am, there is a distinct drop, which is less severe
during the winter months. Moreover, it can be seen that in the
summer, this drop happens earlier than in the other seasons (from
around 6am to 8am). In contrast, during winter time, this drop seems
to happen quite late in the day (from around 9am to 10am), best
visible in Figure \ref{fig_m1}.  This indicates strong interaction
of the wind speed with the sunrise. Over all, the daily mean curves
differ significantly from each other, showing that the diurnal
pattern depends on the annual pattern, and vice versa.

\section{Estimation}\label{section:Estimation}

As an estimation algorithm, we use a lasso based estimation
technique. Lasso is a pe\-na\-lized least square regression method.
Thus, we consider the least squares representation of our model. For
the conditional mean models \eqref{eq_wind_ar_model} and
\eqref{eq_power_ar_model}, the regression representations are given
by

\begin{align}
 \WW_i =& \W_i \bsb^{\WW}_i + \EE_i
\label{eq_wind_ols} \\
 \PP_i =& \P_i \bsb^{\PP}_i + E_i.
\label{eq_power_ols}
\end{align}

Here, $\WW_i = (W_{i, 1}, \ldots, W_{i, n})'$ and $\PP_i = (P_{i, 1}, \ldots, P_{i, n})'$ denote the observed wind speed and wind power vectors, $\W_i$ and $\P_i$ are the matrices of covariates that correspond to \eqref{eq_wind_ar_model} and \eqref{eq_power_ar_model}, $\bsb^{\WW}_i$ and $\bsb^{\PP}_i$ are the full parameter vectors, $\EE_i= (\eps_{i, 1}, \ldots, \eps_{i, n})'$ and $E_i= (\epsilon_{i, 1}, \ldots, \epsilon_{i, n})'$ denote the error vectors and $n$ is the number of observations.\\
Similarly, we formulate a regression representation for the
volatility models. We exploit the fact that $|\eps_{i, t}| =
\gamma_i \sigma_{i, t} + \sigma_{i, t}( |Z_{i, t}| - \gamma_i)$ and
$|\epsilon_{i, t}|^{\frac{1}{3}} = \tau_i
\varsigma^{\frac{1}{3}}_{i, t} + \varsigma_{i, t}^{\frac{1}{3}}(
|U_{i, t}|^{\frac{1}{3}} - \tau_i)$, where $\gamma_i = \E|Z_{i, t}|$
and $\tau_i = \E|Z_{i, t}|^{\frac{1}{3}}$. Note that $v_{i, t} =
\sigma_{i, t}( |Z_{i, t}| - \gamma_i)$ and $u_{i, t} = \varsigma_{i,
t}( |U_{i, t}| - \tau_i)$ are weak white noise processes. With that
we express a recursion on $|\eps_{i, t}|$ and $|\epsilon_{i,
t}|^{\frac{1}{3}}$ by

\begin{align}
|\eps_{i, t}| = &\gamma_i \alpha_{i, 0}(t)  + \sum_{j \in \DD} \sum_{k \in I^{\alpha}_{i, j}} \gamma_i \alpha_{i, j, k}^+(t) \eps^+_{j, t - k} + \gamma_i \alpha_{i, j, k}^-(t) \eps^-_{j, t - k} \nonumber \\
&+ \sum_{j \in \DD} \sum_{k \in I^{\beta}_{i, j}} \gamma_i \beta_{i, j, k}(t) \sigma_{j, t - k} + v_{i, t}
\label{eq_wind_absarch} \\
|\epsilon_{i, t}|^{\frac{1}{3}} =& \tau_i \eta_{i, 0}(t)  + \sum_{j \in \DD} \sum_{k \in I^{\eta}_{i, j}} \tau_i \eta^+_{i, j, k}(t) |\epsilon^+_{j, t - k}|^{\frac{1}{3}} + \eta^-_{i, j, k}(t) |\epsilon^-_{j, t - k}|^{\frac{1}{3}} \nonumber \\
&+ \sum_{j \in \DD} \sum_{k \in I^{\zeta}_{i, j}} \tau_i \zeta_{i, j, k}(t) \varsigma_{j, t - k}^{\frac{1}{3}} \nonumber \\
& + \sum_{j \in \DD} \sum_{k \in I^{\upsilon}_{i, j}} \tau_i
\upsilon^+_{i, j, k}(t) |\eps^+_{j, t - k}|^{\frac{1}{3}}
+ \upsilon^-_{i, j, k}(t) |\eps^-_{j, t - k}|^{\frac{1}{3}} + \nonumber \\
&+ \sum_{j \in \DD} \sum_{k \in I^{\varrho}_{i, j}} \tau_i
\varrho_{i, j, k}(t) \sigma_{j, t - k}^{\frac{1}{3}} + u_{i, t}.
\label{eq_power_absarch}
\end{align}

The corresponding multivariate regression representations are given
by

\begin{align}
|\EE_i| &= \V_i \bsa^{\EE}_i + \VV_i
\label{eq_arch_ols1} \\
|E_i| &= \U_i \bsa^{E}_i + \UU_i,
\label{eq_arch_ols2}
\end{align}

where $\V_i$ and $\U_i$ are the regressor matrices that correspond to \eqref{eq_wind_absarch} and \eqref{eq_power_absarch}, $\bsa_i^{\EE}$ and $\bsa_i^{E}$ are the parameter vectors and $\VV_i = (v_{i,1}, \ldots, v_{i,n})$ and $\UU_i = (u_{i,1}, \ldots, u_{i,n})$.\\
For the parameter estimation we use an estimation technique that is
based on a lasso regression for heteroscedastic data. The approach
is similar to the popular FGLS (feasible generalized least squares)
solution by \cite{Newey1987} at which a weighting matrix brings the
``meat'' into the estimation, leading to point-wise
he\-te\-ro\-sce\-das\-ti\-ci\-ty consistency. This lasso method was analyzed first by \cite{wagener2012bridge} in a standard regression setting and by \cite{ziel2015iteratively} in a time series setting. We slightly modify the algorithm to plug-in the causal setting, i.e. the multivariate approach for the wind power mean model. Therefore, we apply a weighted lasso for the conditional mean regressions \eqref{eq_wind_ols}  and \eqref{eq_power_ols}. For the conditional variance regressions \eqref{eq_arch_ols1} and \eqref{eq_arch_ols2}, we just apply a standard lasso.\\
Let $\bsOmega_i = \text{diag}(\bsomega_i)$ and $\bsXi_i =
\text{diag}(\bsxi_i)$ be diagonal matrices of
he\-te\-ro\-sce\-das\-ti\-ci\-ty weights $\bsomega_i = (\omega_{i,
1},\ldots, \omega_{i, n})$ and $\bsxi_i = (\xi_{i, 1},\ldots,
\xi_{i, n})$. The weighted lasso optimization problems concerning
\eqref{eq_wind_ols}, \eqref{eq_power_ols}, \eqref{eq_arch_ols1} and
\eqref{eq_arch_ols2} are given by

\begin{align}
\what{\bsb}^{\WW}_i &= \argmin_{\bsb } (\WW_i - \W_i \bsb )' \bsOmega  (\WW_i - \W_i \bsb ) + \lambda^{\WW}_{i} |\bsb| \label{eq_lasso_ar_wind} \\
\what{\bsb}^{\PP}_i &= \argmin_{\bsb } (\PP_i - \P_i \bsb )' \bsXi  (\PP_i - \P_i \bsb ) + \lambda^{\PP}_{i} |\bsb| \label{eq_lasso_ar_power} \\
\what{\bsa}^{\EE}_i &= \argmin_{\bsa \geq \bsnull } (|\EE_i| - \U_i \bsa )'  (|\EE_i| - \U_i \bsa ) + \lambda^{\EE}_{i} |\bsa| \label{eq_lasso_arch_wind} \\
\what{\bsa}^{E}_i &= \argmin_{\bsa \geq \bsnull } (|E_i| - \V_i \bsa )'  (|E_i| - \V_i \bsa ) + \lambda^{E}_{i} |\bsa| \label{eq_lasso_arch_power},
\end{align}

with tuning parameters $\lambda_i^{\WW}$, $\lambda_i^{\PP}$, $\lambda_i^{\EE}$ and $\lambda_i^{E}$. Here, we only allow estimators $\what{\bsa}^{\EE}_i$ and $\what{\bsa}^{E}_i$ with no negative entry to ensure that the recurrence equation (and thus, the volatilities $\sigma_{i, t}$ and $\varsigma_{i, t}$) are well defined.\\
For solving the lasso optimization problems we use the coordinate descent algorithm as introduced by \cite{friedman2007pathwise}. This algorithm solves the problem on a given tuning parameter grid. We choose the tuning parameters by the minimalist but rather strict Bayesian information criterion (BIC), to avoid overfitting. Other information criteria or a cross-validation based approach can be also be applied.\\
In the first step, we estimate the conditional mean parameters $\bsb^{\WW}$ and $\bsb^{\PP}$. Then, we consider the estimated residuals for the estimation of the volatility parameters $\bsa^{\EE}$ and $\bsa^{E}$. Afterward, we use the fitted volatilities to redefine the he\-te\-ro\-sce\-das\-ti\-ci\-ty matrices $\bsOmega_i$ and $\bsXi_i$ to repeat the procedure with the new weight matrices for the conditional mean parameters $\bsb^{\WW}$ and $\bsb^{\PP}$. In practice however, there is an initialization problem, as the residuals $\eps_{i, t}$ and $\epsilon_{i, t}$ as well as the conditional standard deviations $\sigma_{i, t}$ and $\varsigma_{i, t}$ are unknown.\\
Still, \cite{chen2011subset} discuss a method for estimating ARMA models using an iterative lasso approach. Their algorithm basically uses the fact that every ARMA($p$, $q$) can be written as an AR($\infty$). Thus, it can be approximated by an AR($p$) for large $p$. This is also applicable for time varying threshold VARMA models. Thus, a time varying threshold VARMA($p$, $q$) process is a time varying threshold AR($\infty$) process. Similarly, for the volatility model, it holds that every exponentiated ARMA model can be expressed as a PGARCH with the corresponding power. For a power of two, a squared ARMA process is a GARCH process. So, the same relationship holds: A time varying power-GARCH($p$, $q$) process is a time varying power-ARCH($\infty$) process.\\
Using these facts we can handle the initialization problem as follows: In the first iteration step, we replace all $\eps_{i, t}$, $\epsilon_{i, t}$, $\sigma_{i, t}$ and $\varsigma_{i, t}$ by $1$. Thus, in the first step, we actually estimate time varying threshold AR processes in the conditional mean equations and time varying power-ARCH processes in the second step.\\
Finally, we initialize the he\-te\-ro\-sce\-das\-ti\-ci\-ty weights $\bsOmega_i$ and $\bsXi_i$. Here, we simply assume homoscedasticity in the first step, so we take $\bsOmega_i = \bsXi_i = I$. The algorithm can be stated as follows:\\

\framebox{
\parbox{0.9\textwidth}{
\begin{centering}
\begin{enumerate}
 \item[\textit{1)}] Initialize $\W_i$, $\P_i$, $\bsOmega_i$, $\bsXi_i$ for all $i\in \DD$ and $K=1$.
 \item[\textit{2)}] Estimate $\what{\bsb}_i^{\WW}$ and $\what{\bsb}_i^{\PP}$ with \eqref{eq_lasso_ar_wind} and \eqref{eq_lasso_ar_power}
 using coordinate descent with weights $\bsOmega_i$ and $\bsXi_i$ for all $i\in \DD$.
 \item[\textit{3)}] Estimate $\what{\bsa}_i^{\EE}$ and $\what{\bsa}_i^{E}$ with \eqref{eq_lasso_arch_wind} and \eqref{eq_lasso_arch_power}
 by coordinate descent using the estimated residuals $(\what{\eps}_{i, 1}, \ldots, \what{\eps}_{i, n} )$ and $(\what{\epsilon}_{i, 1}, \ldots, \what{\epsilon}_{i, n})$
 from \textit{2)} to compute $\U_i$ and $\V_i$ for all $i\in \DD$.
 \item[\textit{4)}] Compute the estimated volatilities $(\what{\sigma}_{i,1}, \ldots, \what{\sigma}_{i,n})$ and $(\what{\varsigma}_{i,1}, \ldots, \what{\varsigma}_{i,n})$ with the fitted values from \textit{3)} and redefine $\W_i$, $\P_i$, $\bsOmega_i = \text{diag}(\bsomega_i)$ and $\bsXi_i = \text{diag}(\bsxi_i)$ by $ \bsomega_i= (\what{\sigma}_{i,1}^{-2}, \ldots, \what{\sigma}_{i, n}^{-2})$ and $ \bsxi_i= (\what{\varsigma}_{i, 1}^{-2}, \ldots, \what{\varsigma}_{i, n}^{-2})$ for all $i \in \DD$.
 \item[\textit{5)}] If $K < K_{\max}$ then $K = K + 1$ and back to \textit{2)}, otherwise stop the algorithm.
 \end{enumerate}
\end{centering}
\vspace{-1mm}
}
}
\vspace{5mm}

We stop the algorithm after a maximum of $K_{\max} = 2$ iterations,
which already provides a good ratio of accuracy and computing time.
\cite{ziel2015iteratively} shows that under some regularity
conditions, two iterations are sufficient to receive optimal
asymptotic properties. As the considered estimation methodology is
based on the coordinate descent algorithm, it shares the same
computational complexity. With $n$ as number of observations, $d$ as
number of turbines and $p$ as dimension of the underlying lasso
problem (dimension of parameter vector in \eqref{eq_power_ols}), the
asymptotic computational complexity of the algorithm is $\OO(d n
p)$. Thus, if either $d$, $n$ or $p$ is doubled, the computation
time gets doubled as well. Notably, the estimation procedure is
easily applicable for large wind parks.

\section{Forecasting and Results}\label{section:Results}

After the estimation, the obtained parameters are fit to the current
set of data in order to calculate a forecast. As it is the very
nature of the lasso approach to return a lot of zero valued
parameters, the high-dimensional parameter space is shrunk to a
manageable amount of relevant parameters, conditional on the unique
settings of in-sample data at each point forecast. We evaluate our
model (``lasso'') and several benchmark approaches according to
their forecasting accuracy. The common criterion is the mean
absolute error (MAE). The out-of-sample (OOS) forecasts are
performed for a time frame from November 2011 to November 2012. Most
benchmark models require appreciable amounts of computing time
(several minutes per forecast). To keep the time consumption
reasonable, we select $N = 1000$ points in time ($\chi^{(l)}, l = 1,
\ldots, N$) in the out-of-sample period at random. For the
respective in-sample periods, we consider the corresponding
preceding year with 52830 observations each. Forecasts are
calculated at horizons of up to a maximum of two days (i.e. 48 hours
= 288 steps). MAE is calculated by

\begin{align}
\text{MAE}_{i, k} &= \frac{1}{N} \sum_{l = 1}^N \left| P_{i,
\chi^{(l)} + k} - \widehat{P}_{i, \tau^{(l)} + k} \right|,
\end{align}

where $\widehat{P}_{i, \chi^{(l)} + k}$ is the $k$-step forecast of
wind power and $P_{i, \chi^{(l)} + k}$ is the corresponding actual
observation, each at station $i$. As the results look similar for
all of the eight turbines, we just report the mean results over all
turbines, so we evaluate

\begin{align}
\text{MAE}_{k} &= \frac{1}{d} \sum_{i = 1}^d \text{MAE}_{i, k}.
\end{align}

Results for the distinct turbines are available upon request.
Additionally to the $\text{MAE}_{k}$ we compute the difference of
$\text{MAE}_{k}$ to the persistent benchmark model, denoted by
$\text{DMAE}_{k}$, i.e.

\begin{align}
\text{DMAE}_{k} &= \text{MAE}_{k} - \text{MAE}^{\text{pers.}}_{k},
\label{eq_dmae}
\end{align}

where $\text{MAE}^{\text{pers.}}_{k}$ is the $\text{MAE}_{k}$ of the
persistent forecaster. The persistent forecaster (so-called na\"\i
ve predictor, $\hat{P}_{\chi^{(l)} + k} = P_{\chi^{(l)}}$, see,
e.g., \citealp{Costa2008}) is suitable to illustrate the improvement
of sophisticated forecasting models in direct comparison to this
common quasi-standard benchmark.\\
We compare our model's results to further benchmarks. We consider a simple univariate AR on the wind power on each turbine $i \in \DD$ (AR), a bivariate VAR on wind power and wind speed of each turbine $i \in \DD$ (BVAR), a $2 \times d = 16-$dimensional multivariate VAR, jointly on all wind power and wind speed processes (abbr.: VAR), the established WPPT model and its recent generalization, \mbox{GWPPT}. Furthermore, we evaluate an ARMA model, an artificial neural network based approach (ANN), and a gradient boosting machine (GBM).\\
The AR-type models (AR, BVAR, VAR) are estimated by solving the
system of Yule-Walker equations, which guarantees a stationary
solution. The corresponding autoregressive order is chosen by
minimizing the Akaike information criterion (AIC). Next to AR-type
models we consider a univariate ARMA(1,1) process benchmark model
for each turbine. \cite{de2011error} states that out of all
ARMA-type models, the ARMA(1,1) process yield the best forecasting
results for a short forecasting horizon. We estimate the ARMA model
by maximizing the Gaussian likelihood.\\
The WPPT is based on a turbine specific dynamic regression approach.
It takes wind speed as a regressor and captures diurnal
pe\-rio\-di\-ci\-ty by a Fourier series of time of day observations
to estimate the parameters of the model

\begin{multline}\label{eq:WPPT}
\hat{P}_{t + k} = m + a_1 \cdot P_{t} + a_2 \cdot P_{t - 1} + b_1
\cdot W_{t + k|t} + b_2 \cdot W_{t + k|t}^2 + d_1^c \cdot
\cos\left(\frac{2\pi d_{t + k}}{144}\right)\\ + d_2^c \cdot
\cos\left(\frac{4\pi d_{t + k}}{144}\right) + d_1^s \cdot
\sin\left(\frac{2\pi d_{t + k}}{144}\right) + d_2^s \cdot
\sin\left(\frac{4\pi d_{t + k}}{144}\right) + \varepsilon_{t + k},
\end{multline}

where $W_{t + k|t}$ is wind speed at time $t + k$ given at time $t$,
$d_t$ is time of day for observation $t$ and $\varepsilon_{t + k}$
is assumed white noise.\\
The generalization of WPPT, \mbox{\mbox{GWPPT}}, is modeled as
both-sided censored: Each wind turbine is manufactured to operate at
a certain range, the so-called power range. \mbox{\mbox{GWPPT}}
makes use of this a-priori known information. The model imposes the
following structure on wind power:

\begin{equation}\label{eq:Censored1}
P_{t}^{*} = \eta(\textbf{z}_{t})+\varepsilon_{t},
\end{equation}

where $\textbf{z}_t$ is the vector of explanatory variables, $\eta$
is a linear function of $\textbf{z}_t$, and $\varepsilon_t$ is an
assumed Gaussian error term. To comply with WPPT,
\mbox{\mbox{GWPPT}} assumes the structure shown in equation
(\ref{eq:WPPT}), but adds wind direction to the specification.
\mbox{\mbox{GWPPT}} imposes a censored data structure, so that

\begin{equation}\label{eq:Censored2}
P_{t} =
  \begin{cases}
  l, ~ &P_{t}^{*}\leq l\\
  P_{t}^{*}, ~ &P_{t}^{*} \in (l,u)\\
  u, ~ &P_{t}^{*}\geq u,
  \end{cases}
\end{equation}

where $l$ and $u$ are the lower and upper censoring points.
Parameters are estimated using a generalized Tobit model. In the
end, due to assumed Gaussian errors, the forecast is calculated by

\begin{equation}\label{eq:CondMean}
\hat{P}_{t + k} = (\Phi(f_2) - \Phi(f_1)) \cdot P_{t + k}^{*} +
(\phi(f_1) - \phi(f_2)) \cdot \widehat{\sigma} + u \cdot (1 -
\Phi(f_2)),
\end{equation}

where $f_1 = (l - P_{t + k}^{*})/\widehat{\sigma}$, $f_2 = (u - P_{t
+ k}^{*})/\widehat{\sigma},$  and $\phi(\cdot)$ and $\Phi(\cdot)$
denote normal PDF (Probability Density Function) and CDF (Cumulative
Distribution Function), respectively.\\
For the ANN benchmark, we consider a single feed-forward neural
network and stick close to the setting as used by
\cite{li2010comparing}. As its inputs, we consider the lagged values
of the past 8 hours of observations. We train the neural network on
50 training vectors and on 4 neurons in the hidden layer.\\
The last benchmark investigated is based on gradient boosting
machines (GBM). \cite{landry2016probabilistic} use GBM methods
successfully for the wind power forecasting track in the Global
Energy Forecasting Competition 2014. We train a GBM for each wind
turbine with a memory of 5 hours on the full data set. Similarly to
\cite{landry2016probabilistic}, we choose the shrinkage tuning
parameter to be 0.05, the interaction depth to be 5 and the minimum
number of observations to be 30. In total, we choose 100 trees,
which are sufficient to reach convergence.

\begin{figure}[h!]
\centering
\begin{subfigure}[b]{0.9\textwidth}
 \includegraphics[width=1\textwidth]{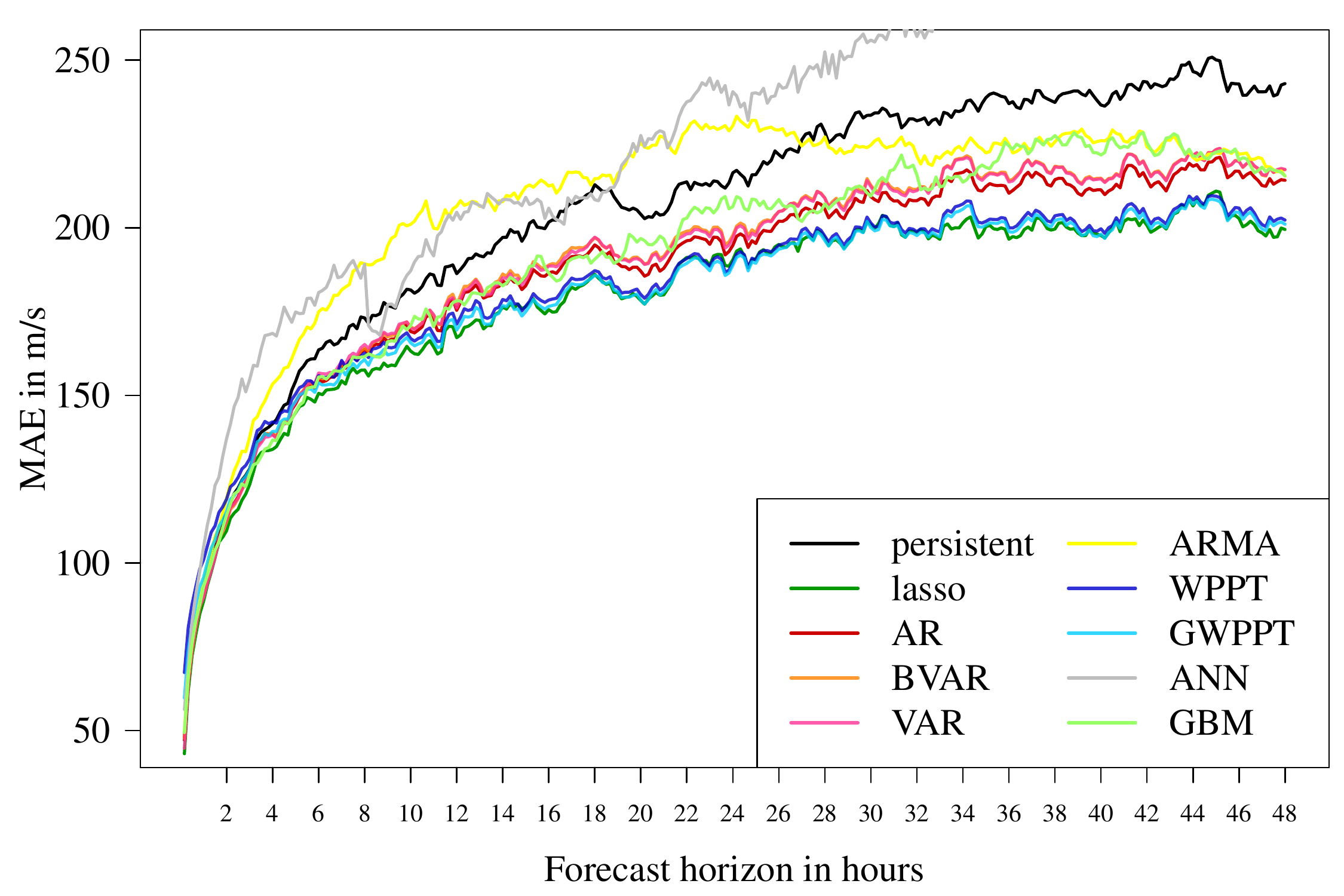}
   \caption{$\text{MAE}_k$ for the selected point forecasts.}
  \label{graph:MAE}
\end{subfigure}
\begin{subfigure}[b]{0.9\textwidth}
 \includegraphics[width=1\textwidth]{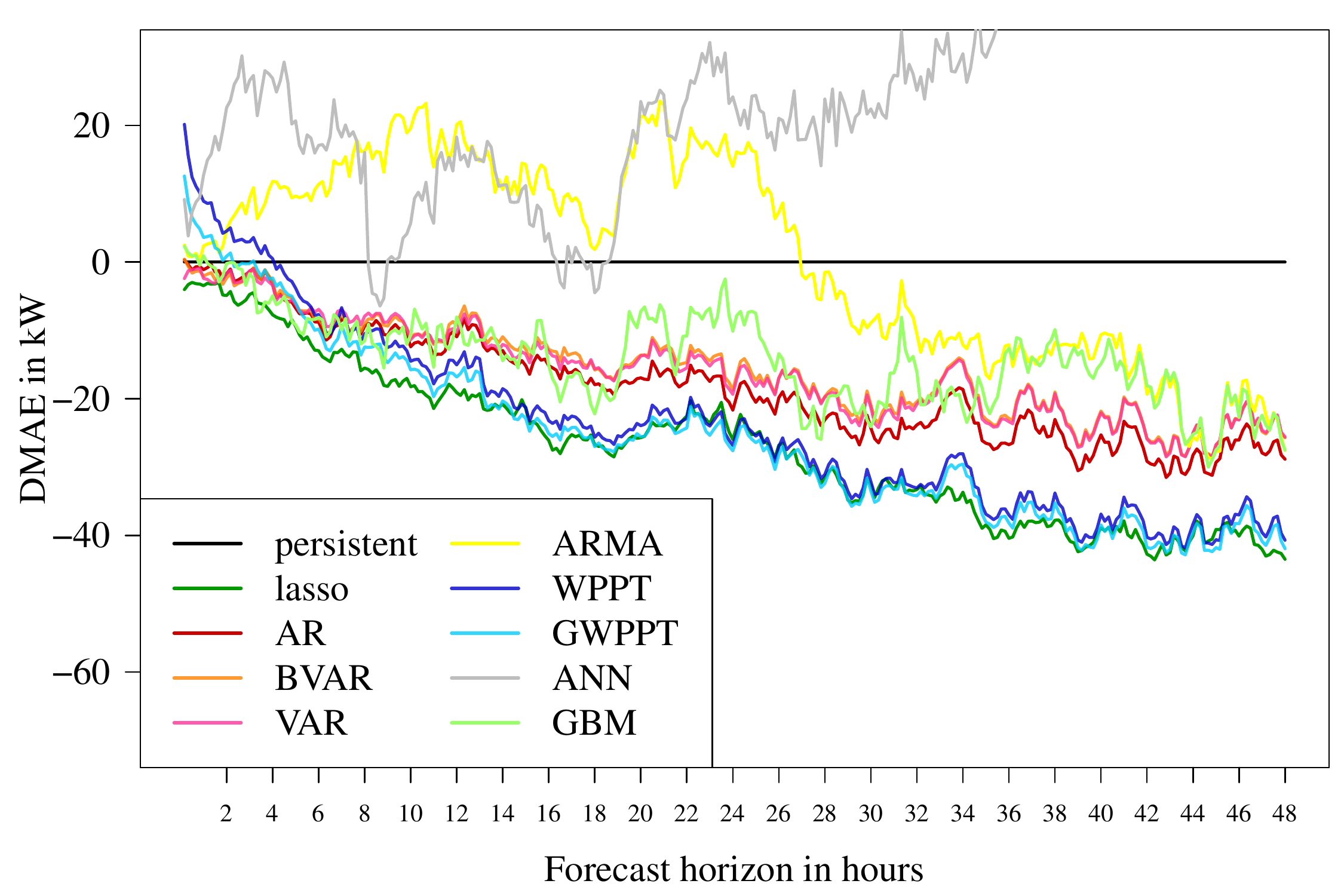}
   \caption{$\text{DMAE}_k$ (difference to persistence) for the selected point forecasts.}
  \label{graph:MAE}
\end{subfigure}
 \caption{$\text{MAE}_k$ and $\text{DMAE}_k$ for all forecasting horizons $k$, time frame from November 2011 to November 2012.}
 \label{graph:Error}
\end{figure}

\begin{figure}[hbt!]
\centering
\begin{subfigure}[b]{0.49\textwidth}
 \includegraphics[width=1\textwidth]{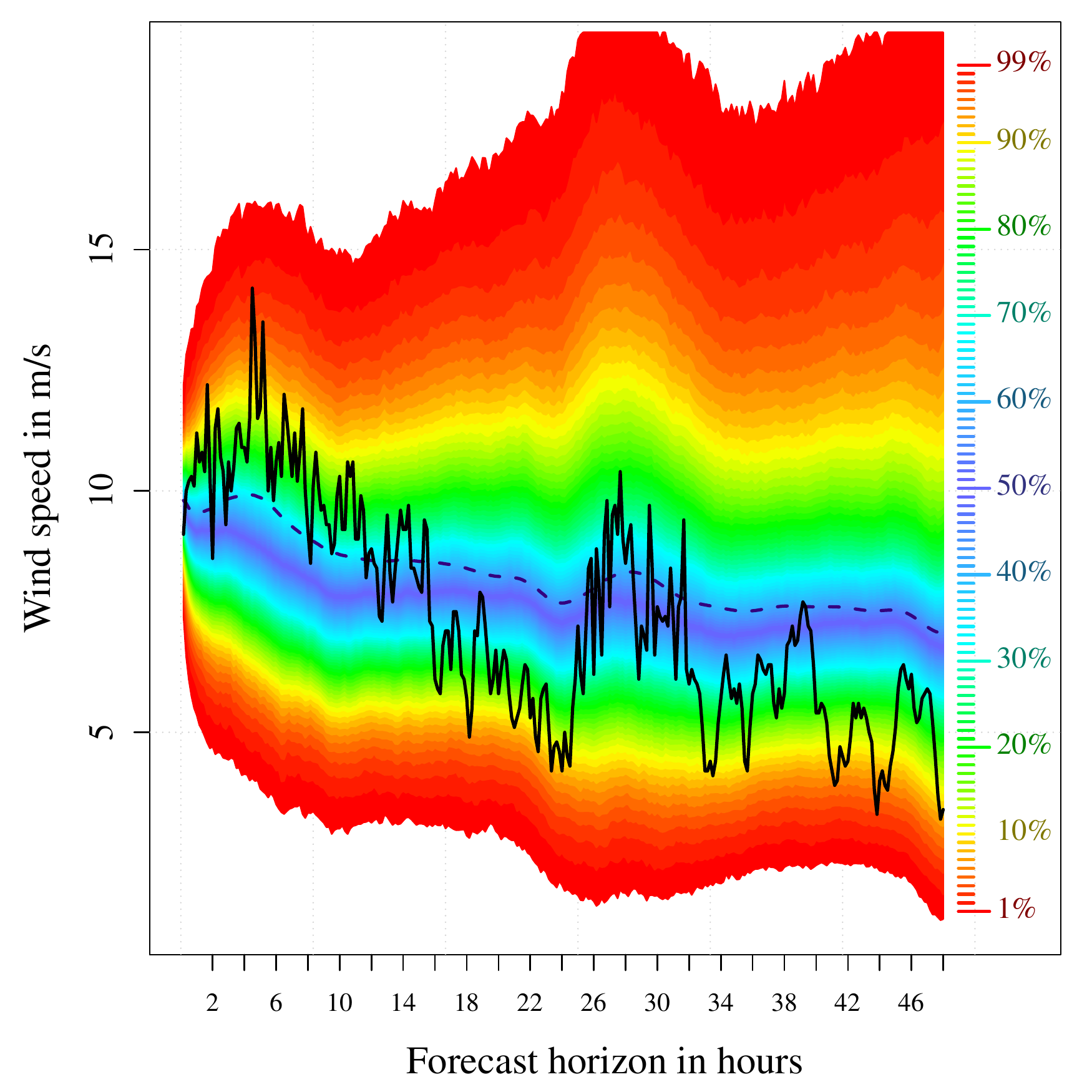}
\caption{Probabilistic wind speed forecast of Turbine A.}
  \label{fig_d1}
\end{subfigure}
\begin{subfigure}[b]{0.49\textwidth}
 \includegraphics[width=1\textwidth]{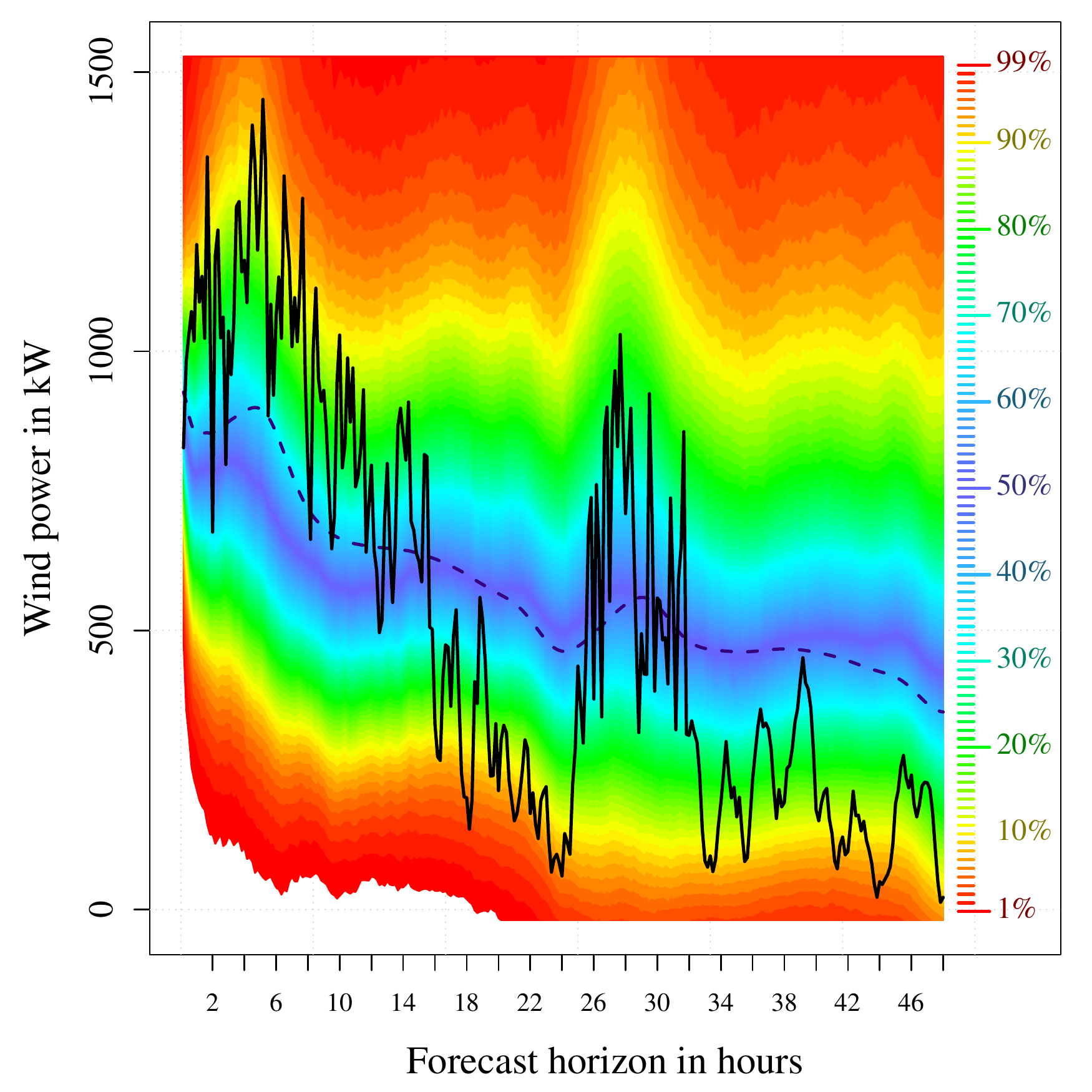}
  \caption{Probabilistic wind power forecast of Turbine A.}
  \label{fig_d2}
\end{subfigure}
\begin{subfigure}[b]{0.49\textwidth}
 \includegraphics[width=1\textwidth]{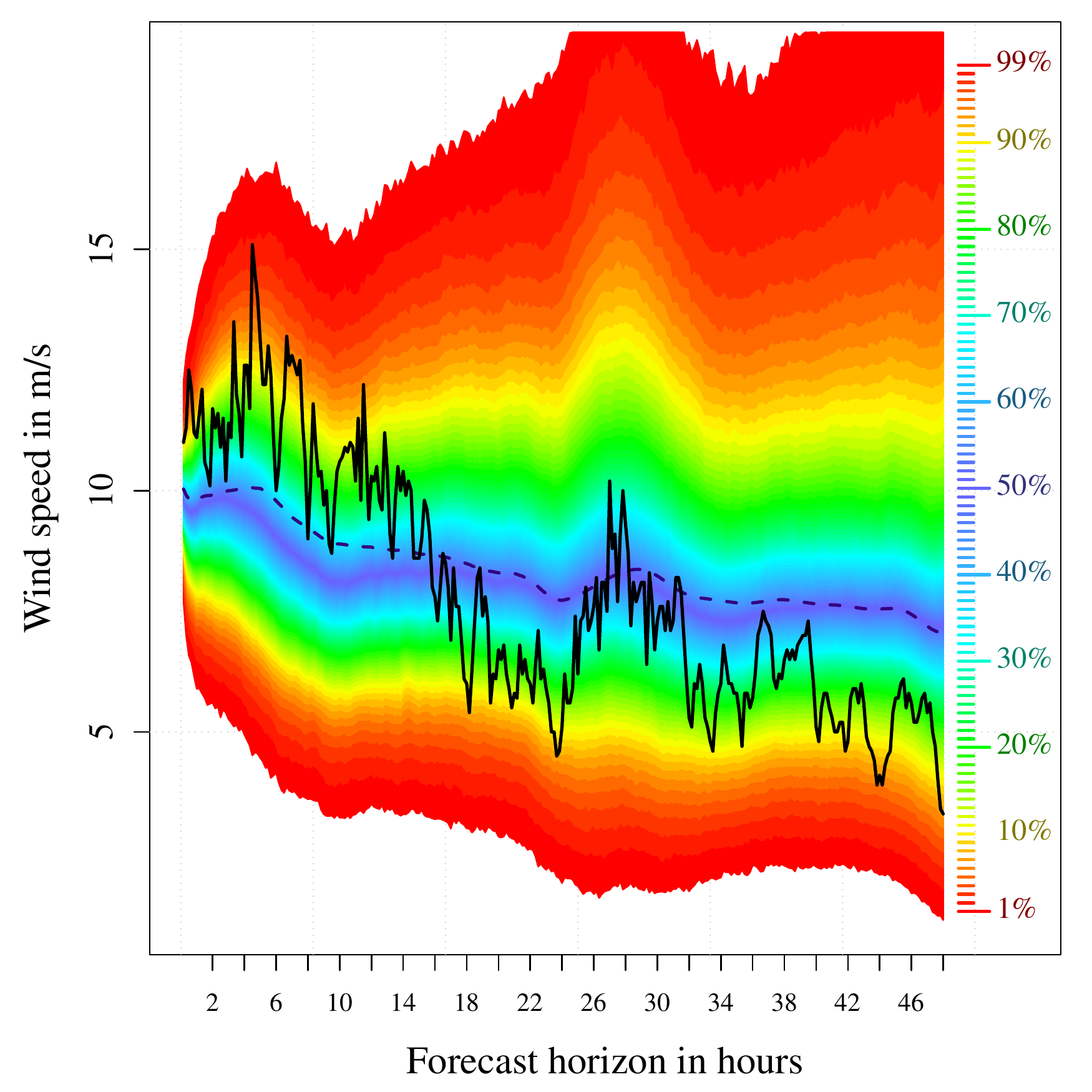}
\caption{Probabilistic wind speed forecast of Turbine B.}
  \label{fig_d1}
\end{subfigure}
\begin{subfigure}[b]{0.49\textwidth}
 \includegraphics[width=1\textwidth]{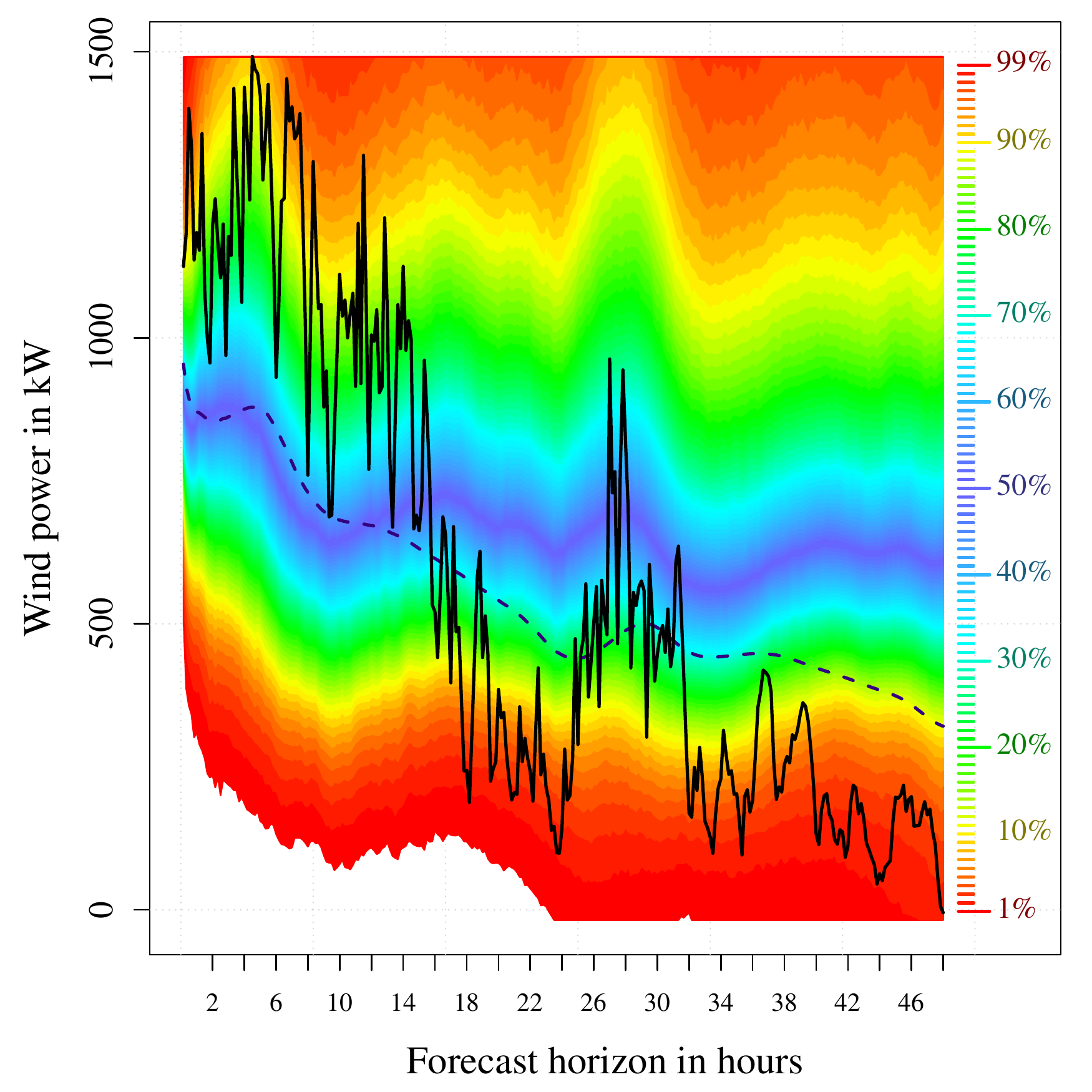}
  \caption{Probabilistic wind power forecast of Turbine B.}
  \label{fig_d2}
\end{subfigure}
 \caption{Probabilistic wind speed and power forecast of Turbines A and B from 2012-02-25 07:20 to 2012-02-27 07:10. The black lines are the observed values, the dashed blue
 lines give the respective point estimates.}
 \label{graph:prob}
\end{figure}

\begin{figure}[hbt!]
\centering
\begin{subfigure}[b]{0.44\textwidth}
 \includegraphics[width=1\textwidth]{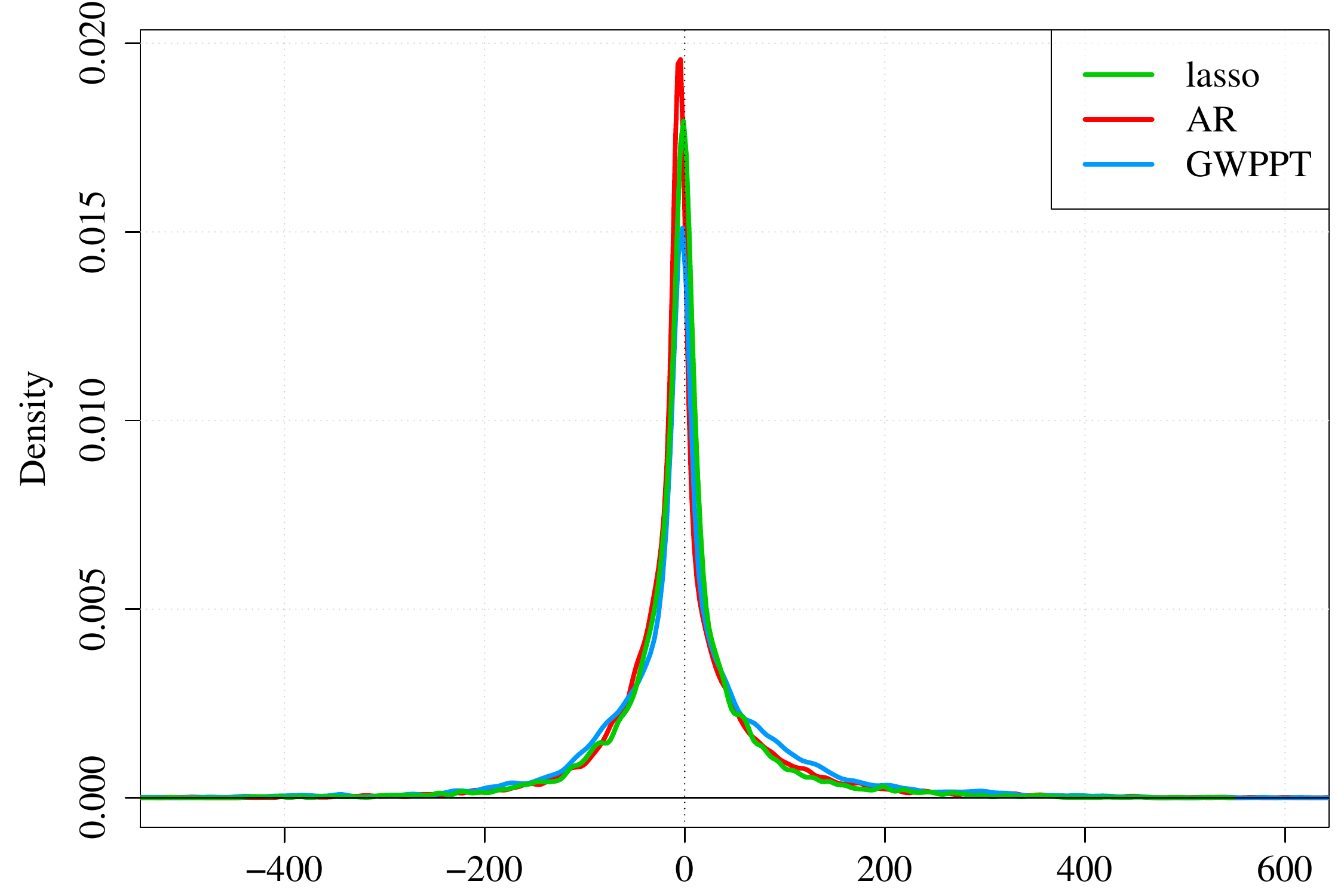}
   \caption{$h=1$}
  \label{graph:Dens1}
\end{subfigure}
\begin{subfigure}[b]{0.44\textwidth}
 \includegraphics[width=1\textwidth]{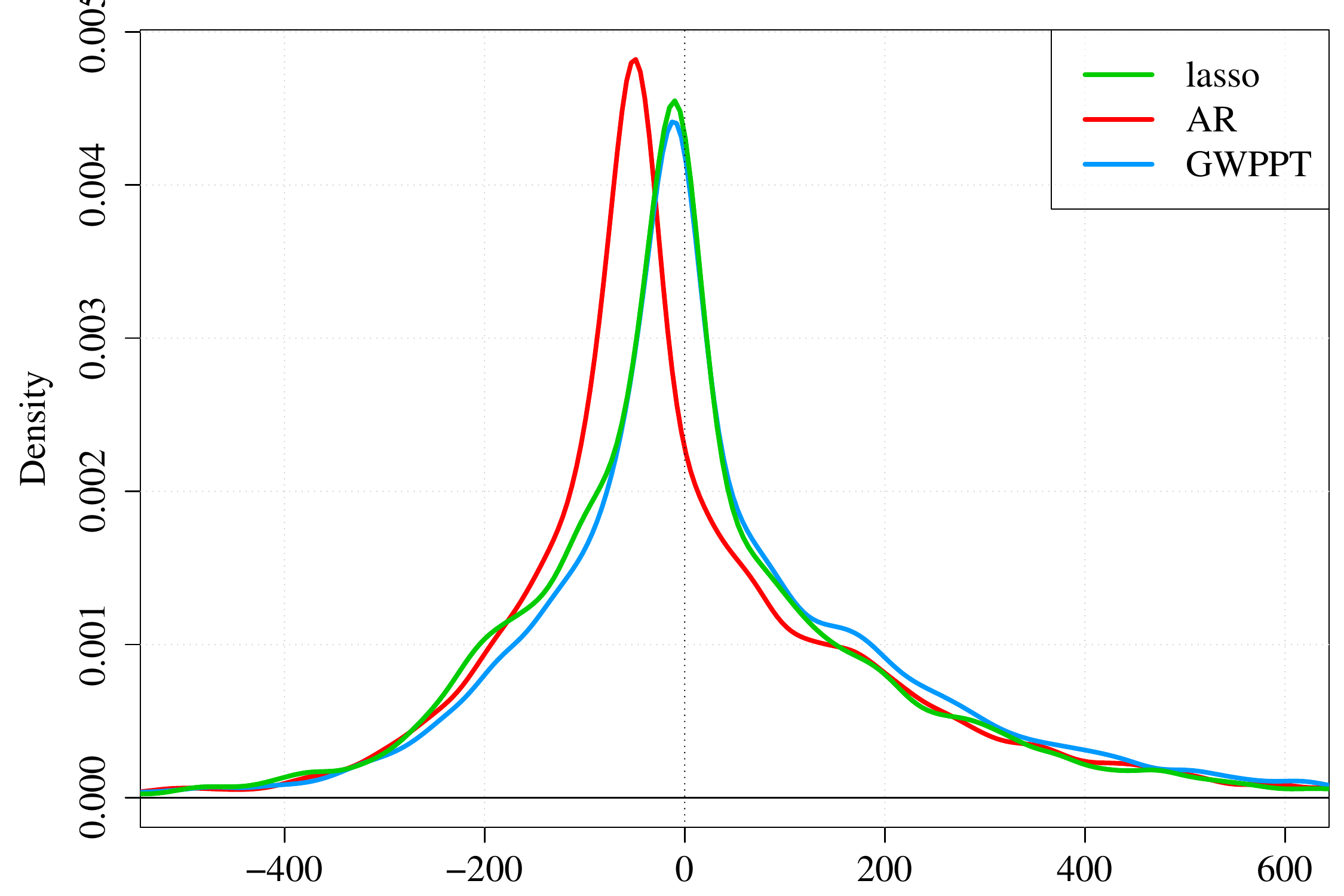}
   \caption{$h=24$}
  \label{graph:Dens24}
\end{subfigure}
\begin{subfigure}[b]{0.44\textwidth}
 \includegraphics[width=1\textwidth]{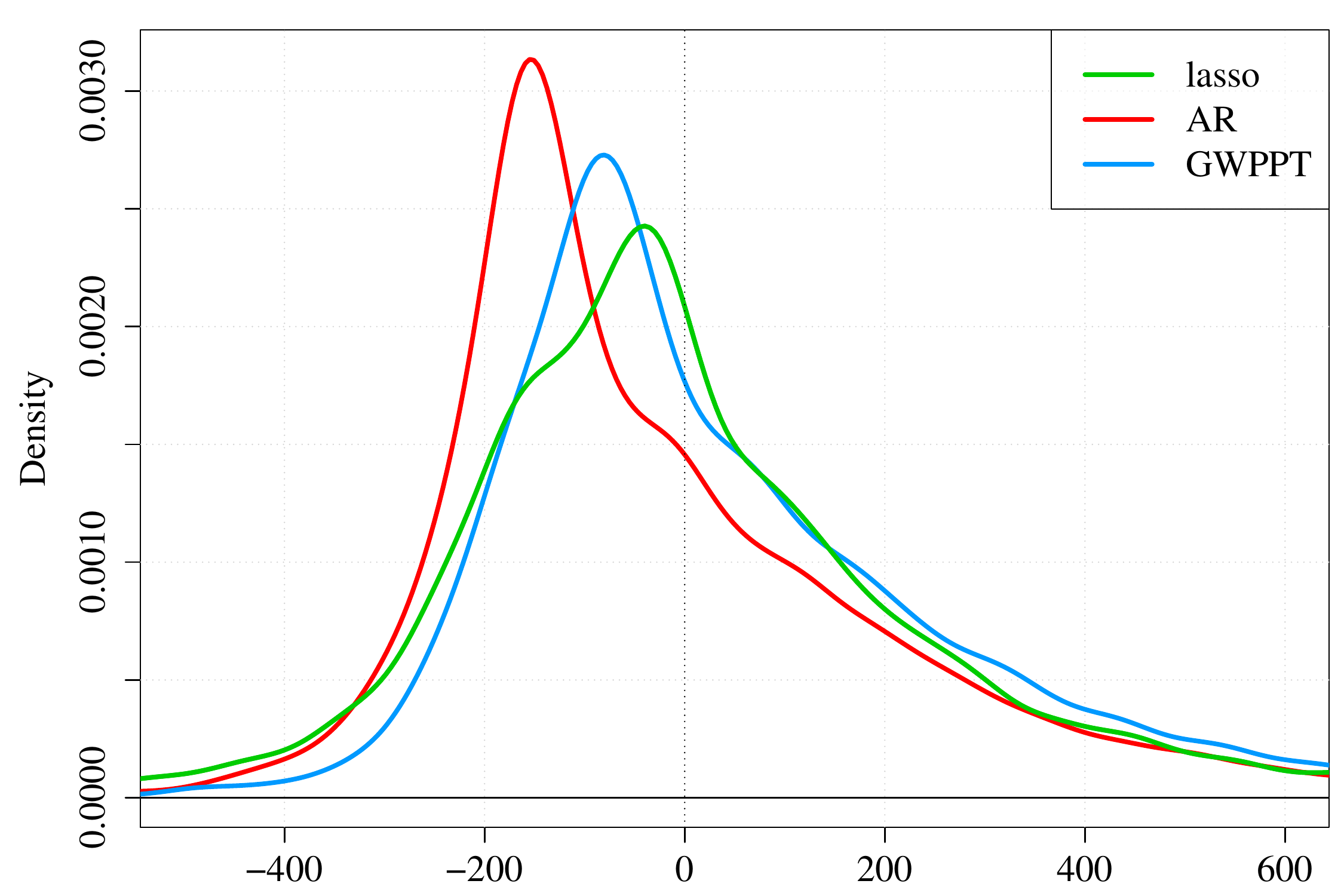}
   \caption{$h=144$}
  \label{graph:Dens144}
\end{subfigure}
\begin{subfigure}[b]{0.44\textwidth}
 \includegraphics[width=1\textwidth]{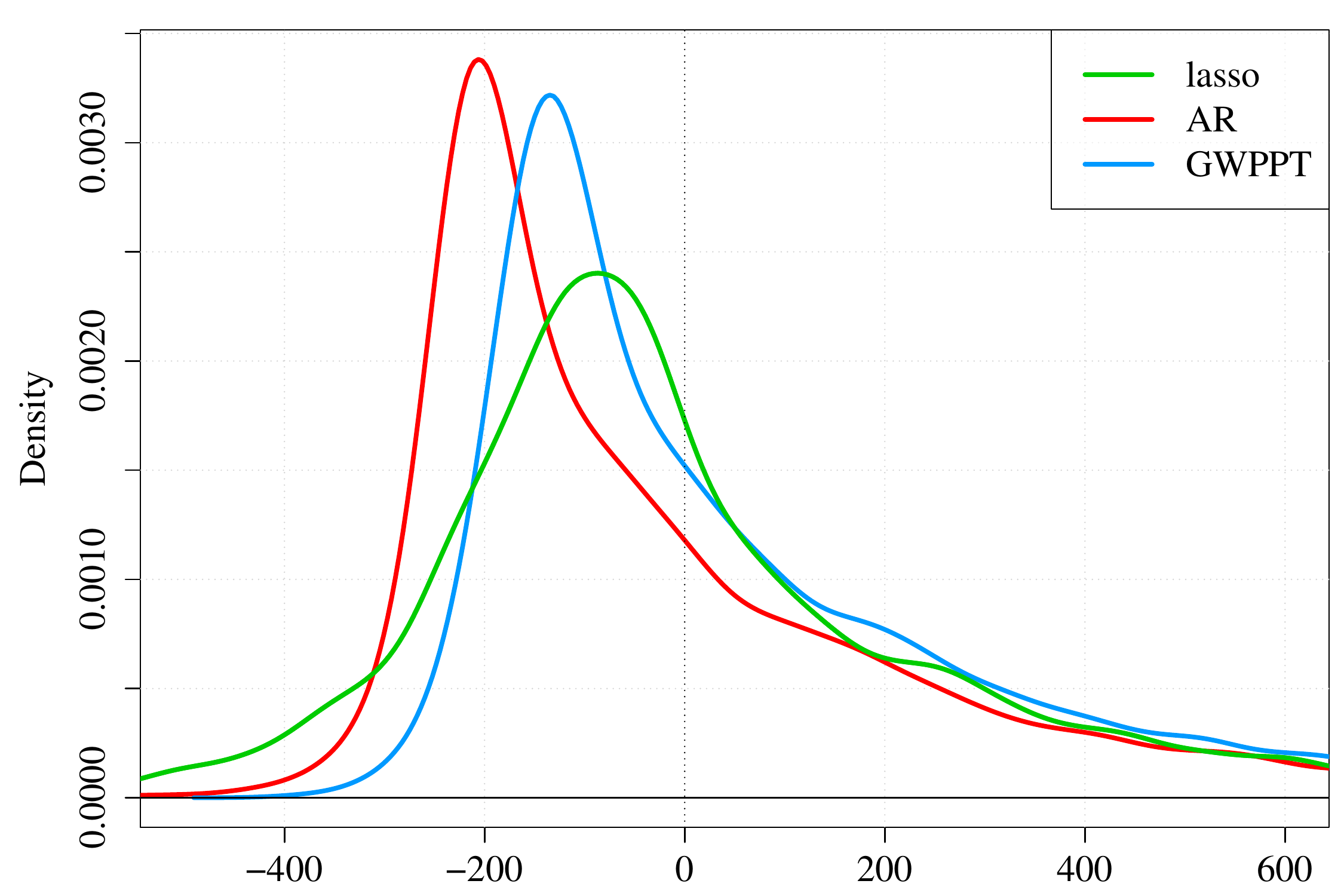}
   \caption{$h=288$}
  \label{graph:Dens288}
\end{subfigure}
 \caption{OOS forecasting errors density, lasso, AR and \mbox{GWPPT}, $h$ step ahead forecasts, time frame from November 2011 to November 2012.}
 \label{graph:Dens}
\end{figure}

Figure \ref{graph:Error}
presents the out-of-sample aggregated forecasting error results. Looking at MAE and DMAE, persistence is outperformed by far by most models. ARMA and ANN perform badly. The GBM and the AR type models perform better, but are still not very competitive. Most of the times, lasso competes with WPPT and \mbox{GWPPT}, but sometimes, lasso outruns (G)WPPT, e.g. at forecasting horizons of around 8 hours and above 32 hours.\\
Table \ref{table:RMSEMAE} shows the results for several selected forecasting horizons (1 step, 6 steps (1 hour), 24 steps (4 hours), 48 steps (8 hours), 72 steps (12 hours), 144 steps (1 day) and 288 steps (2 days)). As can be seen, lasso is either the best model or not significantly different from the best model.\\
Note that for longer forecasting horizons (e.g. 24 or 48 hours), the
surplus of point forecasting is limited due to the strong amount of
uncertainty. However, the proposed model can be used for
probabilistic forecasting as well. Using residuals based bootstrap
as done by, e.g., \cite{ziel2016lasso}, we can easily simulate
sample paths for the wind speed and power of all turbines in a wind
park. We evaluate the empirical quantiles of the bootstrap samples
paths and obtain an estimate for the corresponding quantile.
Exemplarily, Figure \ref{graph:prob} shows the probabilistic wind
speed and power forecast for the 99 percentiles for Turbines A and
B, starting at February 25th, 2012, 07:20. The figure reveals both,
the diurnal seasonal pattern as well as
he\-te\-ro\-sce\-das\-ti\-ci\-ty. For instance, it can be seen that
at a forecasting horizon of $4$ as well as for $24 + 4 = 28$ hours,
there are greater forecasting values for both the wind speed and the
wind power. Indeed, the observations around these peaks are greater
than those in the near proximity. Additionally, these peaks are
rather volatile, so that the prediction intervals at these peaks are
relatively wide, slightly wider than in the neighboring hours.
Overall, we see that the prediction intervals get wider with
increasing forecasting horizon as expected. In general, they seem to
be relatively wide. However, we see that in each of the four figures
some observations fall into the reddish colored area which
represents large prediction intervals. Most distinct, in Figure
\ref{fig_d2} and for large forecasting horizons of more than 40
hours it can be seen that all observations of the wind power of
Turbine B fall into the prediction area of very small probabilities.
Thus, the prediction intervals do
not seem to be too wide or too conservative.\\
Finally, we investigate the OOS errors' asymmetry by looking at the
errors' densities. For more lucidity, we restrain the plots to a few
models, lasso, AR and \mbox{GWPPT}. As Figure \ref{graph:Dens1}
shows, for the one step ahead forecast, all models return symmetric
and leptokurtic results. This symmetry declines for increasing
forecasting horizons. For the 24 steps (4 hours) ahead forecast, AR
starts to tend to asymmetry, as Figure \ref{graph:Dens24} shows. The
average error is negative, which represents a systematic
over-estimation of wind power. \cite{Croonenbroeck2015} show that
this type of bias turns out to be very costly, from a turbine
operator's point of view. \mbox{GWPPT} and the lasso, however, are
still symmetric, mostly. For even longer forecasting horizons
(Figures \ref{graph:Dens144} and \ref{graph:Dens288} show densities
for one day and two days ahead forecasting errors), the AR asymmetry
becomes worse, while \mbox{GWPPT} starts to return asymmetric
forecasts as well. Also, the lasso model becomes asymmetric, but not
as strongly as the other models. From that we conclude that using
the lasso model instead of any of the other models may provide not
only the most accurate forecasts, but also has the least severe
impact of asymmetry, which is important for any turbine operator
with respect to the financial impact of the forecast.

\begin{landscape}
\begin{table}[ht]
\centering
\small
\begin{tabular}{rlllllll}
  \hline
 & 1 & 6 & 24 & 48 & 72 & 144 & 288 \\
  \hline
persistent &  47.12(0.61) &  92.04(1.16) & 141.58(1.59) & 173.09(1.97) & 186.30(2.19) & 216.38(2.42) & 242.99(2.59) \\
  lasso &  \textbf{43.10}(0.52) &  \textbf{88.59}(1.08) & \textbf{133.87}(1.48) & \textbf{157.58}(1.79) & \textbf{167.20}(1.96) & \underline{190.42}(2.19) & \textbf{199.49}(2.12) \\
  AR &  47.45(0.57) &  90.93(1.03) & 138.45(1.34) & 163.73(1.58) & 175.36(1.72) & 194.72(1.88) & 214.14(1.91) \\
  BVAR &  47.55(0.58) &  \underline{89.99}(1.03) & 138.74(1.36) & 164.10(1.58) & 177.08(1.72) & 197.69(1.89) & 217.29(1.94) \\
  VAR &  44.72(0.50) &  \underline{89.64}(1.03) & 138.19(1.35) & 165.18(1.59) & 176.02(1.69) & 197.03(1.87) & 217.40(1.93) \\
  ARMA &  49.61(0.53) &  94.41(0.94) & 153.39(1.44) & 189.23(1.78) & 206.41(1.87) & 230.39(2.10) & 215.40(1.84) \\
  WPPT &  67.30(0.78) & 100.95(1.15) & 142.09(1.52) & 162.59(1.76) & 171.26(1.90) & \underline{189.55}(2.10) & \underline{202.28}(2.17) \\
  \mbox{GWPPT} &  59.72(0.75) &  95.64(1.15) & 139.26(1.52) & \underline{160.66}(1.76) & \underline{169.49}(1.92) & \textbf{188.77}(2.12) & \underline{201.04}(2.21) \\
  ANN &  56.29(0.59) & 104.78(1.22) & 168.50(1.76) & 189.13(2.05) & 204.61(2.06) & 240.57(2.19) & 296.65(2.03) \\
  GBM &  49.42(0.52) &  92.96(0.93) & \underline{136.65}(1.38) & 161.55(1.74) & 178.26(1.79) & 206.27(2.10) & 215.48(2.12) \\
   \hline
\end{tabular}
\caption{MAEs with estimated standard deviations. Best = bold, all
within the 2-sigma range are underlined (not significantly worse
than the best).} \label{table:RMSEMAE}
\end{table}
\end{landscape}

\section{Conclusion}\label{section:Conclusion}
In this paper we present a new wind power forecasting approach that incorporates conditional he\-te\-ro\-sce\-das\-ti\-ci\-ty, flexible periodicity and non-linearity modeling and provides the important wind speed forecasts in only one step. As an estimation technique, we present the re-weighted iterative lasso, which consumes little computing time, provides automatic regularization and sparsity and does not require a distributional assumption, unlike the usual maximum likelihood estimation.\\
The model for wind speed and wind power combines a multivariate time
varying TVARMA process with a power-TGARCH model. The model returns
accurate wind power forecasting results that are competitive,
especially for the medium-term scenario. Furthermore, the model
allows for probabilistic forecasting. Wind park operators may
benefit from the minor asymmetry of our model. While other models
tend to over-estimate in increasing forecasting horizon settings,
our model remains mostly stable, which helps keeping the
asymmetry-induced financial loss of forecasts under control. For
energy markets match-making, finally, our model provides not only
superior accuracy for the point forecast necessary for both sellers
and buyers, but also gives insight into the forecasts' distribution
and thus, the forecasts' reliability.

\clearpage

\appendix
\section{}\label{section:Appendix}
A B-spline basis function of degree $H$ is constructed out of a
B-spline basis function $\wtilde{B}$. $\wtilde{B}$ is defined by the
degree $H$ and a set of knots $\KK$. The set of knot $\KK$ contains
$H + 1$ knots $\{k_0 , \ldots, k_{H + 1}\}$ with $k_h < k_{h + 1}$.
This can be easily defined by the recurrence relation from \cite[p.
90]{deBoor2001book}:

\begin{align}
&\wtilde{B}(t; \{k_0, \ldots, k_{H + 1}\}, H) \nonumber \\
=&  \frac{t - k_0}{k_H - k_0} \wtilde{B}(t; \{k_0, \ldots, k_{H}\}, H - 1) \nonumber \\
&+  \frac{t - k_1}{k_{H + 1} - k_1} \wtilde{B}(t; \{k_1, \ldots, k_{H + 1}\}, H - 1)
\end{align}

with initialization

\[
\wtilde{B}(t; \{k_{l}, k_{l + 1}\}, 0) =
 \begin{cases}
  1 &, t \in [k_l, k_{l + 1}) \\
  0 &, \ow.
 \end{cases}
\]

We consider the set of knots $\KK(T, H)$ to be equidistant with
center $T$. Thus, we find $k_{0} = T - h \frac{D + 1}{2}$, $k_{D +
1} = T + h \frac{D + 1}{2}$ and
since we select an odd degree $D$, we get $k_{\frac{D + 1}{2}} = T$, where $h$ is the distance between the knots. Note that $H$ and $h$ define the knots $\KK$ uniquely.\\
Finally, we consider a seasonality $S$ to obtain a periodic basis function $\wtilde{B}(t; \KK, H)$. To do so, it is suitable to choose $h$ such that $S$ is an integer multiple of $h$, which itself is at least $H + 1$ to guarantee a partition of the unity. We define

\begin{equation}
\wtilde{B}_1^{*}(t; \KK, H) = \sum_{k \in \Z} \wtilde{B}(t - kS; \KK, H)
\end{equation}

as the initial periodic basis function. In our setting, the data has
two seasons, a diurnal and an annual one.\footnote{Wind speed as
well as wind power can be assumed to be periodic for daily and
yearly patterns. Empirically, this behavior can be shown by using
periodograms, i.e. by analyzing the empirical spectral density.} As
our data frequency is at 10 minutes, we have six observations per
hour. Thus, our diurnal seasons are $S_{\text{diurnal}} = 24 \times
6 = 144$ and the yearly seasons are $S_{\text{annual}} = 365.24
\times 24 \times 6 = 52594.56$.\footnote{Note that an average year
lasts 365.242375 days, which is approximated by the leap year system
every four years. A usual consensus is to approximate this by 365.24
days per year.} By using the initial periodic basis function
$\wtilde{B}_1^{*}$, we define the full periodic basis by
$\wtilde{B}^*_j(t; \KK, H)  = \wtilde{B}^*_{j - 1}(t - h; \KK, H)$.
In conclusion, the basis $\BB = \{\wtilde{B}^*_1, \ldots,
\wtilde{B}^*_{N_\BB} \}$
has a total of $N_\BB = S/h$ basis functions. In our setting, we choose $h_{\text{diurnal}} = 12$ and $h_{\text{annual}} = 4$.\\
The basis functions $\wtilde{B}^*_l$ are suitable to capture seasonal changes of parameters. However, due to the structure of the the basis functions, they model the absolute impact over time. In practice, it may be better to consider the changes over time instead of the absolute impact, especially if we use automatic shrinkage and selection algorithms for estimation, just as we do. We can easily model the changes in the parameters over time by cumulating the basis functions
$\wtilde{B}^*_l$ within $l$. Hence, we define

\begin{equation}
\wtilde{B}^{*, \text{cum.}}_l = \wtilde{B}^{*, \text{cum.}}_{l - 1} + \wtilde{B}^{*}_l
\end{equation}

for $l>1$ with $\wtilde{B}^{*, \text{cum.}}_1 = \wtilde{B}^{*}_1$.\\
We use the cumulative basis functions for the conditional mean model \eqref{eq_wind_ar_model}, for both the diurnal and the annual basis functions. Also, we use the non-cumulative version for the conditional variance model \eqref{eq_wind_arch_model}, due to the parameter constraints in the variance model. We discuss this in greater detail in the estimation section.\\
As pointed out by \cite{ziel2015efficient}, there might be interactions between the seasonal components. As the amount of sunshine is changing over the year, this might have impact on the wind speed. Thus, it is possible that daily cyclic effects are changing over the year. The simplest approach to model these interactions is to consider multiplications on the corresponding basis functions. We will use this multiplication for the conditional mean model, where we consider the cumulative basis functions. The multiplications are

\begin{align}
B^{\text{cum.}}_{l_1 h_{\text{annual}} + l_2}(t)
= &\wtilde{B}^{*, \text{cum.}}_{l_1}(t; \KK(h_{\text{annual}}, H), H) \times \nonumber \\
&\wtilde{B}^{*, \text{cum.}}_{l_2}(t; \KK(h_{\text{diurnal}}, H), H)
\end{align}

for $l_1 \in\{1, \ldots, h_{\text{diurnal}}\}$ and $l_2  \in\{1,
\ldots, h_{\text{annual}}\} $ in equation
\eqref{eq_tim-var-coef_basis} and for each periodic coefficient $\xi$ in \eqref{eq_wind_ar_model}.\\
For the conditional variance equation, we do not consider the cumulative basis function, so here the multiplication is

\begin{align}
B_{l_1 h_{\text{annual}} + l_2}(t) =
&\wtilde{B}^{* }_{l_1}(t; \KK(h_{\text{annual}}, H), H) \times \nonumber \\
&\wtilde{B}^{*}_{l_2}(t; \KK(h_{\text{diurnal}}, H), H)
\end{align}

for $l_1 \in\{1, \ldots, h_{\text{diurnal}}\}$ and $l_2  \in\{1,
\ldots, h_{\text{annual}}\} $ in equation
\eqref{eq_tim-var-coef_basis} and for each periodic coefficient $\xi$ of the conditional variance model \eqref{eq_wind_arch_model}.\\
However, by construction of the periodic basis, $\sum_{l=1}^{N_\BB}
\wtilde{B}^*_l (t)$ is constant. Thus, for the time varying
coefficient $\xi$ of the conditional mean model
\eqref{eq_wind_ar_model}, we consider the set of basis functions

\begin{align}
\BB^{\text{cum.}}_{\xi} = \{B^{\text{cum.}}_{l_1 h_{\text{annual}} + l_2} |
&l_1  \in\{1, \ldots, h_{\text{diurnal}}\}, \nonumber \\
& l_2  \in\{1, \ldots, h_{\text{annual}}\}\},
\end{align}

where the last element is constant. Note that $\BB^{\text{cum.}}_{\xi}$ has $h_{\text{diurnal}} \times h_{\text{annual}}$ elements, so that in our setting, $12 \times 4 = 48$ parameters for each time varying coefficient.\\
For the conditional variance model \eqref{eq_wind_arch_model}, we define the used set of basis function for a periodic parameter $\xi$ by

\begin{align}
\BB_{\xi} = \{1\}\cup \{B_{l_1 h_{\text{annual}} + l_2} |
&l_1  \in\{1, \ldots, h_{\text{diurnal}}\}, \nonumber \\
& l_2  \in\{1, \ldots, h_{\text{annual}}\} , \nonumber \\
& (l_1,l_2)\neq (1,1)\}.
\end{align}

Thus, we replace the first basis function $B_1$ by the constant $1$, to model the constant impact directly.

\clearpage

\bibliographystyle{elsarticle-harv}
\bibliography{article}

\end{document}